\documentclass[12pt]{article}
\usepackage[english,activeacute]{babel}
\usepackage{natbib}
\usepackage{comment}
\usepackage{float}
\usepackage[hidelinks]{hyperref}
\usepackage{mathrsfs}
\usepackage{enumitem}
\usepackage[font={small,it}]{caption}
\usepackage{amsmath,amsfonts,amsthm,amssymb}
\usepackage{bm,rotating,multirow,dsfont,graphicx}
\usepackage[usenames, dvipsnames]{color}
\usepackage{url}
\usepackage{multicol}
\usepackage{multirow}
\usepackage[T1]{fontenc}
\usepackage{flafter}
\usepackage{appendix}
\usepackage{subfigure}
\usepackage{xcolor}
\usepackage{soul}
\usepackage{setspace}
\usepackage{booktabs}
\usepackage{tabularx}
\usepackage{setspace}

\makeatletter
\def\hlinewd#1{%
	\noalign{\ifnum0=`}\fi\hrule \@height #1 %
	\futurelet\reserved@a\@xhline}
\makeatother
\def\spacingset#1{\renewcommand{\baselinestretch}{#1}\small\normalsize}
\spacingset{1}

\begin{document}

\title{\textbf{Reyes's \(I\): Measuring Spatial Autocorrelation in Compositions}}

\date{}

\author{
    Lina Buitrago\footnote{Corresponding author: labuitragor@unal.edu.co.}\qquad Juan Sosa\qquad Oscar Melo \\
    \hspace{1cm}\\
    Universidad Nacional de Colombia, Colombia 
}

\maketitle

\begin{abstract}
Compositional observations arise when measurements are recorded as parts of a whole, so that only relative information is meaningful and the natural sample space is the simplex equipped with Aitchison geometry. Despite extensive development of compositional methods, a direct analogue of Moran's \(I\) for assessing spatial autocorrelation in areal compositional data has been lacking. We propose Reyes's \(I\), a Moran type statistic defined through the Aitchison inner product and norm, which is invariant to scale, to permutations of the parts, and to the choice of the \(\operatorname{ilr}\) contrast matrix. Under the randomization assumption, we derive an upper bound, the expected value, and the noncentral second moment, and we describe exact and Monte Carlo permutation procedures for inference. Through simulations covering identical, independent, and spatially correlated compositions under multiple covariance structures and neighborhood definitions, we show that Reyes's \(I\) provides stable behavior, competitive calibration, and improved efficiency relative to a naive alternative based on averaging componentwise Moran statistics. We illustrate practical utility by studying the spatial dependence of a composition measuring COVID-19 severity across Colombian departments during January 2021, documenting significant positive autocorrelation early in the month that attenuates over time.
\end{abstract}


\spacingset{1.1} % DON'T change the spacing!

\newpage

\section{Introduction}

Compositional data analysis (CoDa) emerges when handling data that represent proportions or components of a whole.
Applying traditional multivariate statistical methods to analyze such data leads to erroneous results due to spurious correlations, as the value of each component inherently depends on the values of the others.
In this context, \cite{aitchison1982}, \cite{barcelo2001}, \cite{pawlowsky_egozcue2001geo}, \cite{egozcue2003}, and \cite{aitchison_egozcue2005} make substantial contributions by introducing key concepts such as the sample space (the simplex), Aitchison geometry in the simplex as a Euclidean space, the principle of working on coordinates, distributions within the simplex, and linear models involving compositional response or explanatory variables.
When working with the simplex, compositional data are typically treated as vectors with a constant sum (usually 1). However, \cite{barcelo2001} presents compositions as equivalence classes, lifting the constant-sum constraint. In this work, we adopt the former approach and apply Aitchison geometry within the simplex.
\cite{pawlowsky_glahn2011} and \cite{egozcue2019compositional} provide a valuable overview of the theory and practical applications of CoDa.

CoDa has been extensively applied across various fields, with medicine being one of the most relevant examples.
For instance, \cite{mandal2015microb}, \cite{lin2020microb}, \cite{gacesa2022microb}, and \cite{nearing2022microb} utilize CoDa to analyze the gut microbiome. 
Additionally, \cite{fernandes2014genomics} apply CoDa to RNA sequencing analysis, while \cite{dumuid2018act} and \cite{zhao2021dietary} use it to investigate lifestyle patterning.
Furthermore, \cite{hernandez2022machine} and \cite{hedou2024biomark} incorporate machine learning and deep learning techniques into their analyses, with the latter specifically developing methods to identify omics biomarkers.

Compositional spatial data represent another significant source of applications in which CoDa has gained considerable popularity, primarily focusing on geostatistical data.
This type of data analysis has been analyzed using alternative methods, such as Multinomial models with Dirichlet priors distributions \citep{tjelmeland2003bayesian,pirzamanbein2018bayesian}, along with machine learning approaches for spatial interpolation \citep{nwaila2024spatial}.
Nonetheless, \cite{pawlowsky2016review} presents several advancements in the analysis of geostatistical data, primarily focused on the use of isometric log-ratio (ilr) coordinates \citep{paw2015modeling}.

To the best of our knowledge, however, spatial autocorrelation indicators for compositional data have not been proposed yet, similar to those established by \cite{moran1950} for real-valued data \citep[e.g.,][]{yamada2024new}. 
Such indicator is crucial for both spatial and compositional literature because, akin to Moran's I for spatial data, it enables the assessment of whether neighboring areas exhibit similar values. 
In this article, we propose a spatial autocorrelation indicator specifically designed for georeferenced compositional data, thoroughly examining its properties through both theoretical analysis and empirical investigation using synthetic data.

Our proposal possesses both practical and statistical utility, as it enables decision-making grounded in spatial behavior of compositions.
For example, it can be very useful when carrying out model selection tasks.
If spatial autocorrelation is absent and the composition is analyzed as a response variable, a traditional linear model may be sufficient. Conversely, if spatial autocorrelation is present, it is essential to adjust the model to account for this relationship.

An alternative to our proposal would be to calculate Moran's I for each component of the composition and summarize the resulting values using a preferred summary statistic, such as the mean.
However, we strongly prevent the reader from such an approach, as it fails to account for the interdependence between the components, which could potentially lead to misleading results and, consequently, erroneous decision-making. 
In our simulation study, we explore this naive alternative and demonstrate how it can be highly harmful under certain conditions.

The analysis of disease severity distribution is a critical issue that can be effectively addressed using compositional data, since the vector representing the proportions of patients in different severity states constitutes a composition.
Determining the presence of spatial autocorrelation in this composition is essential for making informed public health decisions.
For instance, if the spatial autocorrelation is positive (i.e., neighboring regions exhibit similar compositional patterns), it is advisable to implement consistent prevention or intervention measures across neighboring regions, such as lockdowns and travel restrictions, especially in the case of infectious diseases whose transmission vectors spread spatially.

In this article, we specifically analyze COVID-19 in Colombia, by defining a composition based on the patient's place of care (home, hospital, or intensive care unit), as a proxy for disease severity. 
Thus, we assess whether spatial autocorrelation exists during the month of January 2021. 
Had this analysis been conducted earlier in the pandemic, Colombian authorities could have made more informed decisions regarding disease control as well as vaccination.

The structure of the manuscript is organized as follows: 
Section 2 introduces the analysis of compositional data.
Section 3 describes the fundamental concepts of spatial statistics for data in \(\mathbb{R}\). 
Section 4 presents a detailed theoretical study of the Reyes's I, an indicator specifically designed to quantify spatial autocorrelation in compositional data.
Section 5 applies the proposed indicator to analyze COVID-19 cases in Colombia.
Section 6 presents a comprehensive simulation study to empirically evaluate the properties of the proposed indicator. 
Finally, Section 7 discusses the main findings and outlines potential directions for future research.

\section{Compositional Data Analysis}

A vector $\boldsymbol{x} = (x_1, \ldots, x_D)$ is said to be a composition if $x_j > 0$, for $j = 1, \ldots, D$, and its components carry only \textit{relative information}. In other words, the relevant information is contained solely in the ratios between the components \citep{pawlowsky_glahn2011}. When subject to a constant sum constraint, the sample space of compositions is the simplex \citep{paw2015modeling}:
$$
\mathcal{S}^D = \{ (x_1, \ldots, x_D) \in (\mathbb{R}^{+})^{D} : \textstyle\sum_{i=1}^D x_i = k \}\,,
$$
with $k \in \mathbb{R}$ a fixed constant. Given this constraint, Euclidean geometry is not suitable for analysis, as the value of one component necessarily depends on the values of the others, leading to spurious correlations between components. Therefore, for proper analysis, Aitchison geometry on the simplex must be used \citep{pawlowsky_egozcue2001geo}.

\subsection{Aitchison Geometry}

Given $\boldsymbol{x} = (x_1, \ldots, x_D) \in \mathcal{S}^D$, $\boldsymbol{y} = (y_1, \ldots, y_D) \in \mathcal{S}^D$, and $\alpha \in \mathbb{R}$, here, we follow the work of \cite{paw2015modeling} to define key operations for carrying out CoDa, including perturbation, powering, and the Aitchison inner product.

Perturbation corresponds to the addition of compositions. Formally, the perturbation of $\boldsymbol{x}$ by $\boldsymbol{y}$ is given by:
\begin{equation*}
    \boldsymbol{x}\oplus\boldsymbol{y} = \mathcal{C}\left(x_1y_1, x_2y_2, \dots, x_Dy_D\right),
\end{equation*}
where $\mathcal{C}(.)$ is the closure operator
$$
\mathcal{C}(\boldsymbol{x}) = \left(\frac{k\,x_1}{\sum_{i=1}^D x_i}, \frac{k\,x_2}{\sum_{i=1}^D x_i}, \dots, \frac{k\,x_D}{\sum_{i=1}^D x_i}\right),
$$
and $k \in \mathbb{R}$ is fixed constant typically set to 1.
Similarly, the inverse operation, analogous to subtraction, is defined as:
\begin{equation*}
    \boldsymbol{x}\ominus\boldsymbol{y} = \mathcal{C}\left(\frac{x_1}{y_1}, \frac{x_2}{y_2}, \dots, \frac{x_D}{y_D}\right).
\end{equation*}

Additionally, powering refers to raising the parts of a composition to a scalar value. It is useful for studying changes in the relative importance of the components under different rescaling conditions. The operation is defined given by:
$$
\alpha\odot\boldsymbol{x} = \mathcal{C}\left(x_1^\alpha,x_2^\alpha,\ldots,x_D^\alpha\right). 
$$

On the other hand, in order to work effectively within a metric space, a distance must be defined, which in turn requires specifying an inner product. For the simplex, the Aitchison inner product is defined as
\begin{equation}\label{prodp}
    \left\langle\boldsymbol{x},\boldsymbol{y}\right\rangle_a = \frac{1}{2D} \sum_{i=1}^D \sum_{j=1}^D \ln \frac{x_i}{x_j} \ln \frac{y_i}{y_j}.
\end{equation} 
Therefore, the Aitchison norm is given by
\begin{equation}\label{norma}
    \left\|\boldsymbol{x}\right\|_a = \sqrt{\left\langle\boldsymbol{x}, \boldsymbol{x}\right\rangle_a} = \sqrt{\frac{1}{2D} \sum_{i=1}^D \sum_{j=1}^D \ln \left( \frac{x_i}{x_j} \right)^2}.
\end{equation}

It can be shown that the simplex, equipped with the Aitchison inner product and the powering operation as an external product, forms a vector space \citep{paw2015modeling}. Consequently, a composition can be expressed as a linear combination of a basis, and in particular, an orthonormal basis. 
This fact enables working with coordinates, allowing us to compute $\mathbb{R}^{D-1}$ coordinates for a composition in $\mathcal{S}^D$, analyze them (typically using classical methods), and then apply an inverse transformation to map the results back to $\mathcal{S}^D$.
The most commonly used coordinates (also known as transformations) in CoDa are the additive log-ratio (\textsf{alr}), centered log-ratio (\textsf{clr}), and isometric log-ratio (\textsf{ilr}), with the latter being the only one based on an orthonormal basis. In this work, we use the \textsf{clr} and \textsf{ilr} coordinates.

\subsection{Centered Log-Ratio Transformation: \textsf{clr}}

Given $\boldsymbol{x} \in \mathcal{S}^D$, the \textsf{clr} coordinates of $\boldsymbol{x}$ are defined as follows \citep{aitchison1982}:
\begin{equation}\label{clr}
    \operatorname{clr}\left(\boldsymbol{x}\right)=\left(\ln\frac{x_1}{g_m(\boldsymbol{x})}, \ln\frac{x_2}{g_m(\boldsymbol{x})}, \dots, \ln\frac{x_D}{g_m(\boldsymbol{x})}\right)\,,
\end{equation}
where 
$$
g_m(\boldsymbol{x}) = \left(\prod_{j=1}^D x_j\right)^{1/D}
$$
is the geometric mean of $\boldsymbol{x}$.
This transformation is an isometry between $\mathbb{R}^{D-1}$ and $\mathcal{S}^D$, meaning that it preserves distances \citep{paw2015modeling}. 
Indeed, for any \(\boldsymbol{x}, \boldsymbol{y} \in \mathcal{S}^D\), the following holds:
$$
\left\langle \boldsymbol{x}, \boldsymbol{y} \right\rangle_a = \left\langle \operatorname{clr}\left(\boldsymbol{x}\right), \operatorname{clr}\left(\boldsymbol{y}\right) \right\rangle,
$$
where \(\left\langle \cdot, \cdot \right\rangle\) represents the inner product in \(\mathbb{R}\). Moreover, we have \(\left\langle \operatorname{clr}\left(\boldsymbol{x}\right), \boldsymbol{1}_D \right\rangle = 0\), where \(\boldsymbol{1}_D\) is a row vector of length \(D\) with all entries equal to one.

\subsection{Isometric Log-Ratio Transformation: \textsf{ilr}}  
\label{subsec:ilr_section}

In contrast to \textsf{clr} coordinates, which correspond to an oblique basis of the simplex, \textsf{ilr} coordinates are associated with an orthonormal basis. 
Formally, let \(\left\{\boldsymbol{e}_1, \boldsymbol{e}_2, \dots, \boldsymbol{e}_{D-1}\right\}\) be an orthonormal basis for \(\mathcal{S}^D\). 
The \textsf{ilr} coordinates are defined as follows \citep{egozcue2003}:
\begin{equation}\label{ilr}
    \operatorname{ilr}\left(\boldsymbol{x}\right) = \operatorname{clr}\left(\boldsymbol{x}\right)\Psi^\textsf{T} = \ln\left(\boldsymbol{x}\right)\Psi^\textsf{T}\,,
\end{equation}
where \(\Psi\) is a contrast matrix of size \(D \times (D-1)\), with the \(i\)-th row given by \(\Psi_i = \operatorname{clr}\left(\boldsymbol{e}_i\right)\).  
Furthermore, for \(\boldsymbol{x}, \boldsymbol{y} \in \mathcal{S}^D\) and \(\alpha, \beta \in \mathbb{R}\), the following properties hold \citep{paw2015modeling}:
\begin{enumerate}
    \item \(\operatorname{ilr}\left(\alpha \odot \boldsymbol{x} \oplus \beta \odot \boldsymbol{y}\right) = \alpha \operatorname{ilr}\left(\boldsymbol{x}\right) + \beta \operatorname{ilr}\left(\boldsymbol{y}\right)\).
    \item \(\left\langle \boldsymbol{x}, \boldsymbol{y}\right\rangle_a = \left\langle \operatorname{ilr}\left(\boldsymbol{x}\right), \operatorname{ilr}\left(\boldsymbol{y}\right)\right\rangle\), i.e., \(\operatorname{ilr}\) is an isometry between \(\mathcal{S}^D\) and \(\mathbb{R}^{D-1}\).
    \item \(\Psi \Psi^\textsf{T} = (\mathbf{I}_D - \boldsymbol{1}_D^\textsf{T}\boldsymbol{1}_D)\), where \(\mathbf{I}_D\) is the identity matrix of size \(D \times D\).
\end{enumerate}
Lastly, note that the \(\operatorname{ilr}\) transformation is a function with domain in the simplex and codomain in \(\mathbb{R}^{D-1}\), whose inverse is giben by:
$$
\operatorname{ilr}^{-1}\left(\boldsymbol{x}^*\right) = \mathcal{C}\left(\exp\left(\boldsymbol{x}^* \Psi\right)\right)\,,
$$
with $\boldsymbol{x}^*\in\mathbb{R}^{D-1}$.

\section{Spatial Statistics for Real-Valued Data}

Let \(X\) be a random variable (r.v.) measured across subareas, and let \(x_i\) represent the observed value of \(X\) in subarea \(i\), for \(i = 1, \ldots, n\).

\subsection{Spatial Autocorrelation}\label{subsection:autocorr}

Spatial autocorrelation exists when the observed values \(x_1,\ldots,x_n\) exhibit interdependence across space \citep{cliff_ord1981}. One of the most commonly used measures to determine the presence of spatial autocorrelation is Moran's \(I\) \citep{moran1950, cliff_ord1981, anselin1995local}:
\begin{equation}\label{I}
    I = n\,\frac{\sum_{i=1}^{n} z_i \Tilde{z}_i}{S_0 \sum_{i=1}^{n} z_i^2}\,,
\end{equation}
where \(z_i = x_i - \bar{x}\), \(\Tilde{z}_i = \sum_{j=1}^{n} w_{ij} z_j\) is the spatial lag of \(z_i\), \(S_0 = \sum_{i=1}^{n} \sum_{j=1}^{n} w_{ij}\), and \(w_{ij}\) represents the \((i,j)\) entry of the spatial weight matrix \(\mathbf{W} = [w_{ij}]\).

The moments of \(I\) are assessed under various assumptions. In this article, we will adopt the \textit{assumption of randomization}.
Under this assumption, the population distribution is not considered. Instead, the observed value of \(I\) is regarded as one of the numerous potential values that could be achieved through all possible permutations of \(x_1, \ldots, x_n\) within the system of regions \citep{cliff_ord1981}.

Equation \eqref{I} clearly illustrates that the value of Moran's index depends on the neighborhood structure, which is represented by the spatial weight matrix \(\mathbf{W}\).
Depending on the problem at hand, the researcher can establish the criteria for defining neighborhoods, typically based on either the distances between the centroids of the subareas or their contiguity. 
Among these criteria are the rook and queen methods. In the rook method, neighboring subareas are defined as those that can be accessed by moving in the same manner as a rook in chess. In contrast, the queen method considers all subareas reachable by moving like a queen to be neighbors.
Thus, the entry \((i,j)\) of \(\mathbf{W}\) is defined as follows:
\begin{equation*}
    w_{ij} = \left\{ \begin{array}{ll} c_i & \text{if } i \text{ is a neighbor of } j; \\ 0 & \text{otherwise,} \end{array} \right.
\end{equation*}
where \(c_i > 0\). Typically, the matrix \(\mathbf{W}\) is row-standardized, ensuring that \(\sum_{j=1}^n w_{ij} = m_i c_i = 1\), where \(m_i\) denotes the number of neighbors of \(i\).

\section{A Spatial Autocorrelation Indicator for Compositional Data}

Traditional spatial autocorrelation measures, like the Moran's I, are typically used to analyze geographically dependent real-valued data. 
However, these measures may not be suitable for compositional data. 
In this section, we introduce the Reyes's I, a compositional adaptation of Moran's I, designed to account for the structure of compositional data using Aitchison geometry. 
The proposed indicator provides a robust framework for analyzing spatial dependencies in compositional datasets.

\subsection{Definition}

Let $\boldsymbol{x}_1, \dots, \boldsymbol{x}_n \in S^D$ be the observed values of the composition across $n$ polygons and $\mathbf{W}$ the spatial weight matrix. 
Building on Equation \eqref{I}, the Reyes's I (compositional adaptation of Moran’s I) is defined as:
\begin{equation}\label{defIa}
I_a = n \sum_{i=1}^{n} \frac{\left\langle \boldsymbol{z}_i, \Tilde{\boldsymbol{z}}_i \right\rangle_a}{S_0 \sum_{i=1}^{n} ||\boldsymbol{z}_k||_a^2}\,,
\end{equation}
where $S_0 = \sum_{i=1}^{n} \sum_{j=1}^{n} w_{ij}$, and $<\cdot,\cdot>_a$ and $\|\cdot\|_a$ correspond to the Aitchison inner product and norm, respectively. 
Additionally, $\boldsymbol{z}_i = \boldsymbol{x}_i \ominus \hat{\boldsymbol{g}}$, for $i = 1,\ldots,n$, with $\hat{\boldsymbol{g}} = \mathcal{C}(\hat{g}_1, \dots, \hat{g}_D)$ and $\hat{g}_j = \left( \prod_{i=1}^{n} x_{ij} \right)^{1/n}$, for $j=1,\ldots,D$. 
Furthermore, $\Tilde{\mathbf{Z}} = \mathbf{W}\Delta\mathbf{Z} = \operatorname{ilr}^{-1}(\mathbf{W}\operatorname{ilr}(\mathbf{Z})) = (\Tilde{\boldsymbol{z}}_1^\textsf{T}, \dots, \Tilde{\boldsymbol{z}}_n^\textsf{T})^\textsf{T}$, where $\Tilde{\boldsymbol{z}}_i$ is the spatial lag of $\boldsymbol{z}_i$.

As of now, without loss of generality, we assume that the spatial weights are row-standardized, meaning that $\sum_{j=1}^{n} w_{ij} = 1$. 
Therefore, $S_0 = \sum_{i=1}^{n} \sum_{j=1}^{n} w_{ij} = n$.

\subsection{Properties}

Let $\boldsymbol{x}_1, \dots, \boldsymbol{x}_n \in S^D$ be the observed compositional vectors over \(n\) polygons, and let \(\mathbf{W}\) denote the spatial weight matrix. The following properties make \(I_a\) well defined and practically useful for measuring compositional spatial autocorrelation. Invariance to the choice of the \(\operatorname{ilr}\) contrast matrix ensures that \(I_a\) depends only on the Aitchison geometry of the compositions and not on arbitrary coordinate representations, which is essential for interpretability and reproducibility. The upper bound controls the magnitude of the statistic and helps identify extreme values or numerical instabilities. Finally, the first and second randomization moments provide key ingredients to standardize \(I_a\) and to construct hypothesis tests or approximations to its null distribution, turning a descriptive autocorrelation measure into an inferential tool that is comparable across datasets, weight matrices \(\mathbf{W}\), and spatial resolutions.

\textbf{Result 1} $I_a$ can be obtained using the $\textsf{ilr}$ transformation and it is invariant with respect to the choice of the contrast matrix $\Psi$.

\textit{Proof:} Following Section \ref{subsec:ilr_section}, we have that:
\begin{align*}
    I_a = \sum_{i=1}^{n} \frac{\left\langle \boldsymbol{z}_i, \Tilde{\boldsymbol{z}}_i \right\rangle_a}{\sum_{i=1}^{n} \|\boldsymbol{z}_k\|_a^2}
    = \sum_{i=1}^{n} \frac{\left\langle \operatorname{ilr}\left(\boldsymbol{z}_i\right), \operatorname{ilr}\left(\Tilde{\boldsymbol{z}}_i\right) \right\rangle}{\sum_{i=1}^{n} \|\operatorname{ilr}\left(\boldsymbol{z}_k\right)\|^2}\,.
\end{align*}

Furthermore, based on Equation \eqref{ilr}:
\begin{align*}
    \left\langle \operatorname{ilr}\left(\boldsymbol{z}_i\right), \operatorname{ilr}\left(\Tilde{\boldsymbol{z}}_i\right) \right\rangle &= \operatorname{ilr}\left(\boldsymbol{z}_i\right) \operatorname{ilr}^\textsf{T}\left(\boldsymbol{z}_i\right) \\
    &= \operatorname{clr}\left(\boldsymbol{z}_i\right) \Psi^\textsf{T} \Psi \operatorname{clr}^\textsf{T}\left(\boldsymbol{z}_i\right) \\
    &= \operatorname{clr}\left(\boldsymbol{z}_i\right)(I_D - \boldsymbol{1}^\textsf{T}_D \boldsymbol{1}_D) \operatorname{clr}^\textsf{T}\left(\boldsymbol{z}_i\right) \\
    &= \operatorname{clr}\left(\boldsymbol{z}_i\right) \operatorname{clr}^\textsf{T}\left(\boldsymbol{z}_i\right) - \operatorname{clr}\left(\boldsymbol{z}_i\right) \boldsymbol{1}^\textsf{T}_D \boldsymbol{1}_D \operatorname{clr}^\textsf{T}\left(\boldsymbol{z}_i\right) \\
    &= \operatorname{clr}\left(\boldsymbol{z}_i\right) \operatorname{clr}^\textsf{T}\left(\boldsymbol{z}_i\right)
\end{align*}

Since $\operatorname{clr}\left(\boldsymbol{z}_i\right) \boldsymbol{1}^\textsf{T}_D \boldsymbol{1}_D = 0$. Additionally, since $\|\boldsymbol{z}_k\|_a^2 = \left\langle \boldsymbol{z}_k, \Tilde{\boldsymbol{z}}_k \right\rangle_a$, it follows that $I_a$ does not depend on the contrast matrix. \hfill$\blacksquare$

\textbf{Result 2} An upper bound for $|I_a|$ is
\begin{equation}
    |I_a|\leq\frac{\sum_{i=1}^{n}||\boldsymbol{z}_i||_a||\Tilde{\boldsymbol{z}}_i||_a}{\sum_{k=1}^{n}||\boldsymbol{z}_k||_a^2}.
\end{equation}\label{bound}

\textit{Proof:}
By the Cauchy--Schwarz inequality,
$$\big|\left\langle\boldsymbol{z}_i,\Tilde{\boldsymbol{z}}_i\right\rangle_a\big|\leq ||\boldsymbol{z}_i||_a\,||\Tilde{\boldsymbol{z}}_i||_a.$$

Since $\sum_{k=1}^{n}||\boldsymbol{z}_k||_a^2\geq 0$,
$$\frac{\big|\left\langle\boldsymbol{z}_i,\Tilde{\boldsymbol{z}}_i\right\rangle_a\big|}{\sum_{k=1}^{n}||\boldsymbol{z}_k||_a^2}\leq\frac{||\boldsymbol{z}_i||_a\,||\Tilde{\boldsymbol{z}}_i||_a}{\sum_{k=1}^{n}||\boldsymbol{z}_k||_a^2}.$$

On the other hand, by the triangle inequality,
$\big|\sum_{i=1}^{n}\left\langle\boldsymbol{z}_i,\Tilde{\boldsymbol{z}}_i\right\rangle_a\big|\leq\sum_{i=1}^{n}\big|\left\langle\boldsymbol{z}_i,\Tilde{\boldsymbol{z}}_i\right\rangle_a\big|$, hence
\begin{equation*}
    |I_a|\leq\frac{\sum_{i=1}^{n}\big|\left\langle\boldsymbol{z}_i,\Tilde{\boldsymbol{z}}_i\right\rangle_a\big|}{\sum_{k=1}^{n}||\boldsymbol{z}_k||_a^2}\leq\frac{\sum_{i=1}^{n}||\boldsymbol{z}_i||_a\,||\Tilde{\boldsymbol{z}}_i||_a}{\sum_{k=1}^{n}||\boldsymbol{z}_k||_a^2}.
\end{equation*}
\hfill$\blacksquare$

Under the randomization assumption, also called the permutation or random labeling assumption, the observed compositional vectors are treated as fixed, and the only source of randomness is the random assignment of these vectors to the \(n\) polygons, while the spatial structure encoded by \(\mathbf{W}\) is held fixed. This assumption is important because it defines a principled reference distribution for \(I_a\) without additional parametric assumptions, supports exact or Monte Carlo permutation tests, and provides analytic benchmark moments used to center and calibrate inference.

\textbf{Result 3:} Under the randomization assumption, the first moment is given by
\begin{equation}
    \textsf{E}_{\text{R}}(I_a)=-\frac{1}{n-1}.
\end{equation}

\textit{Proof:} Let $\textsf{E}_{\text{R}}(||\boldsymbol{Z}_i||^2_a)=\frac{1}{n}\sum_{k=1}^{n}||\boldsymbol{z}_k||_a^2=m$. On the other hand,
\begin{equation}
    \operatorname{ilr}(\Tilde{\mathbf{Z}})=\mathbf{W}\operatorname{ilr}(\mathbf{Z})
    =\left(\begin{matrix}
        \operatorname{ilr}(\Tilde{\boldsymbol{z}}_1)\\
        \vdots\\
        \operatorname{ilr}(\Tilde{\boldsymbol{z}}_n)
    \end{matrix}\right)
    =\left(\begin{matrix}
        \sum_{j=1}^{n}w_{1j}\operatorname{ilr}(\boldsymbol{z}_j)\\
        \vdots\\
        \sum_{j=1}^{n}w_{nj}\operatorname{ilr}(\boldsymbol{z}_j)
    \end{matrix}\right).
\end{equation}

Likewise, by Property 2 in Section~\ref{subsec:ilr_section},
\begin{equation}\label{puntozrz}
 \left\langle\boldsymbol{Z}_i,\Tilde{\boldsymbol{Z}}_i\right\rangle_a
 =\left\langle\operatorname{ilr}(\boldsymbol{Z}_i), \operatorname{ilr}(\Tilde{\boldsymbol{Z}}_i)\right\rangle
 =\sum_{j=1}^{n}w_{ij}\operatorname{ilr}(\boldsymbol{Z}_i)^\top \operatorname{ilr}(\boldsymbol{Z}_j),
\end{equation}
with $i\neq j$. Hence, the expected value of $\left\langle\boldsymbol{Z}_i,\Tilde{\boldsymbol{Z}}_i\right\rangle_a$ under the randomization assumption is given by
\begin{align*}
    \textsf{E}_{\text{R}}\left(\left\langle\boldsymbol{Z}_i,\Tilde{\boldsymbol{Z}}_i\right\rangle_a\right)
    &=\textsf{E}_{\text{R}}\left[\left\langle\operatorname{ilr}(\boldsymbol{Z}_i), \operatorname{ilr}(\Tilde{\boldsymbol{Z}}_i)\right\rangle\right]\\
    &=\textsf{E}_{\text{R}}\left[\textsf{E}_{\text{R}}\left[\left\langle\operatorname{ilr}(\boldsymbol{Z}_i), \operatorname{ilr}(\Tilde{\boldsymbol{Z}}_i)\right\rangle \mid \operatorname{ilr}(\boldsymbol{Z}_i)\right]\right]\\
    &=\textsf{E}_{\text{R}}\left[\textsf{E}_{\text{R}}\left[\sum_{j=1}^{n}w_{ij}\operatorname{ilr}(\boldsymbol{Z}_i)^\top \operatorname{ilr}(\boldsymbol{Z}_j)\mid \operatorname{ilr}(\boldsymbol{Z}_i)\right]\right]\\
    &=\textsf{E}_{\text{R}}\left[\sum_{j=1}^{n} w_{ij}\operatorname{ilr}\left(\boldsymbol{Z}_{i}\right)^\top
    \frac{1}{n-1} \sum_{\substack{l=1\\l\neq i}}^{n} \operatorname{ilr}\left(\boldsymbol{Z}_{l}\right)\right] \\
    &=\textsf{E}_{\text{R}}\left[\sum_{j=1}^{n} \frac{w_{ij}}{n-1}\operatorname{ilr}\left(\boldsymbol{Z}_{i}\right)^\top
    \left(\sum_{l=1}^{n} \operatorname{ilr}\left(\boldsymbol{Z}_{l}\right)-\operatorname{ilr}\left(\boldsymbol{Z}_{i}\right)\right)\right].
\end{align*}

Since $\sum_{l=1}^{n} \operatorname{ilr}\left(\boldsymbol{Z}_{l}\right)=0$, because the compositions $\boldsymbol{Z}_{l}$, $l=1,\ldots,n$, are centered, we obtain
\begin{align*}
\textsf{E}_{\text{R}}\left[\left\langle \boldsymbol{Z}_i, \tilde{\boldsymbol{Z}}_i\right\rangle_a\right]
&=-\frac{1}{n-1} \sum_{j=1}^n w_{i j}\,
\textsf{E}_{\text{R}}\left[\operatorname{ilr}\left(\boldsymbol{Z}_i\right)^\top \operatorname{ilr}\left(\boldsymbol{Z}_i\right)\right] \\
&=-\frac{1}{n-1} \sum_{j=1}^n w_{i j}\left\|\boldsymbol{Z}_i\right\|_a^2
=-\frac{m}{n-1}.
\end{align*}

Therefore,
\begin{equation}\label{esperanza}
    \textsf{E}_{\text{R}}\left(I_a\right)
    =\textsf{E}_{\text{R}}\left[\sum_{i=1}^n \frac{\left\langle\boldsymbol{Z}_i, \tilde{\boldsymbol{Z}}_i\right\rangle_a}{\sum_{k=1}^n\left\|\boldsymbol{Z}_k\right\|_a^2}\right]
    =\frac{\sum_{i=1}^n \textsf{E}_{\text{R}}\left(\left\langle \boldsymbol{Z}_i, \tilde{\boldsymbol{Z}}_i\right\rangle_a\right)}{n m}
    =-\frac{1}{n-1}.
\end{equation}
\hfill$\blacksquare$

\textbf{Result 4} Under the randomization assumption, the second moment is given by
\begin{equation}
    \textsf{E}_{\text{R}}\left(I_a^2\right)
    =\textsf{E}_{\text{R}}\left[\left(\frac{\sum_{i=1}^n\left\langle\boldsymbol{Z}_i, \tilde{\boldsymbol{Z}}_i\right\rangle_a}{\sum_{k=1}^n\left\|\boldsymbol{Z}_k\right\|_a^2}\right)^2\right]
    =\frac{\textsf{E}_{\text{R}}\left[\left(\sum_{i=1}^n\left\langle\boldsymbol{Z}_i, \tilde{\boldsymbol{Z}}_i\right\rangle_a\right)^2\right]}{n^2 m^2}.
\end{equation}\label{momento2}

\textit{Proof:} To derive the second moment of the compositional Moran statistic under randomization, we work with the distribution induced by permuting the observed compositions across the \(n\) spatial units while keeping the spatial weights fixed. Under this randomization measure, the denominator of \(I_a\) is constant, so the problem reduces to computing the second moment of the numerator.

\begin{equation}\label{eq:ER_Ia2_decomp}
\textsf{E}_{\text{R}}\!\left(I_a^2\right)
=
\textsf{E}_{\text{R}}\!\left[\left(\frac{\sum_{i=1}^n\left\langle\boldsymbol{Z}_i,\widetilde{\boldsymbol{Z}}_i\right\rangle_a}{\sum_{k=1}^n\left\|\boldsymbol{Z}_k\right\|_a^2}\right)^2\right]
=
\frac{\textsf{E}_{\text{R}}\!\left(S^2\right)}{(nm)^2},
\quad
S=\sum_{i=1}^n\left\langle\boldsymbol{Z}_i,\widetilde{\boldsymbol{Z}}_i\right\rangle_a,
\quad
m=\frac{1}{n}\sum_{k=1}^n\left\|\boldsymbol{Z}_k\right\|_a^2 .
\end{equation}

Let \(\boldsymbol{z}_i=\operatorname{ilr}(\boldsymbol{Z}_i)\in\mathbb{R}^{D-1}\). With row standardized weights \(w_{ij}\) satisfying \(w_{ii}=0\) and \(\sum_{j=1}^n w_{ij}=1\), define
\begin{equation}\label{eq:Ai_def}
A_i
=
\left\langle\boldsymbol{Z}_i,\widetilde{\boldsymbol{Z}}_i\right\rangle_a
=
\sum_{\substack{j=1\\ j\neq i}}^n w_{ij}\,\boldsymbol{z}_i^\top \boldsymbol{z}_j,
\qquad
S=\sum_{i=1}^n A_i,
\qquad
S^2=\sum_{i=1}^n A_i^2+\sum_{\substack{i=1\\ i\neq j}}^n\sum_{j=1}^n A_iA_j .
\end{equation}

Introduce the empirical second moment matrix and the associated scalars
\begin{equation}\label{eq:M2_m_m4}
\mathbf{M}_2=\frac{1}{n}\sum_{l=1}^n \boldsymbol{z}_l\boldsymbol{z}_l^\top,
\qquad
m=\operatorname{tr}(\mathbf{M}_2),
\qquad
m_4=\frac{1}{n}\sum_{l=1}^n \left(\boldsymbol{z}_l^\top \boldsymbol{z}_l\right)^2 .
\end{equation}

Under randomization over locations, for distinct indices \(i\neq j\) and \(i\neq j\neq k\),
\begin{equation}\label{eq:key_randomization_moments}
\textsf{E}_{\text{R}}\!\left[\left(\boldsymbol{z}_i^\top \boldsymbol{z}_j\right)^2\right]
=\frac{n\,\operatorname{tr}\!\left(\mathbf{M}_2^2\right)-m_4}{n-1},
\qquad
\textsf{E}_{\text{R}}\!\left[\left(\boldsymbol{z}_i^\top \boldsymbol{z}_j\right)\left(\boldsymbol{z}_i^\top \boldsymbol{z}_k\right)\right]
=\frac{2m_4-n\,\operatorname{tr}\!\left(\mathbf{M}_2^2\right)}{(n-1)(n-2)} .
\end{equation}

Let \(c_i=\sum_{j=1}^n w_{ij}^2\). Using \eqref{eq:key_randomization_moments} in the expansion of \(A_i^2\) gives
\begin{equation}\label{eq:ER_Ai2}
\textsf{E}_{\text{R}}(A_i^2)
=
\frac{n\,\operatorname{tr}\!\left(\mathbf{M}_2^2\right)-m_4}{n-1}\,c_i
+
\frac{2m_4-n\,\operatorname{tr}\!\left(\mathbf{M}_2^2\right)}{(n-1)(n-2)}\,(1-c_i).
\end{equation}

For \(i\neq j\), let \(c_{ij}=\sum_{k=1}^n w_{ik}w_{jk}\). Separating the terms in \(A_iA_j\) according to whether they involve distinct indices or a shared index yields the compact form
\begin{equation}\label{eq:ER_AiAj}
\textsf{E}_{\text{R}}(A_iA_j)
=
\frac{2n\,\operatorname{tr}\!\left(\mathbf{M}_2^2\right)+n m^2-6m_4}{(n-1)(n-2)(n-3)}\,(1-c_{ij}-w_{ji})
+
\frac{2m_4-n\,\operatorname{tr}\!\left(\mathbf{M}_2^2\right)}{(n-1)(n-2)}\,c_{ij},
\qquad i\neq j.
\end{equation}

Combining \eqref{eq:ER_Ia2_decomp}, \eqref{eq:Ai_def}, \eqref{eq:ER_Ai2}, and \eqref{eq:ER_AiAj} yields
\begin{equation}\label{eq:ER_Ia2_final}
\textsf{E}_{\text{R}}\!\left(I_a^2\right)
=
\frac{1}{n^2m^2}
\left\{
\sum_{i=1}^n \textsf{E}_{\text{R}}(A_i^2)
+
\sum_{\substack{i=1\\ i\neq j}}^n\sum_{j=1}^n \textsf{E}_{\text{R}}(A_iA_j)
\right\},
\end{equation}
where \(\textsf{E}_{\text{R}}(A_i^2)\) and \(\textsf{E}_{\text{R}}(A_iA_j)\) are given in \eqref{eq:ER_Ai2} and \eqref{eq:ER_AiAj}, respectively, and \(m\) is defined in \eqref{eq:M2_m_m4}. \hfill$\blacksquare$

\subsection{Distribution of \texorpdfstring{$I_a$}{Ia} under the randomization assumption}\label{subsec:distIa}

The distribution of $I_a$ can be computed exactly by enumerating all permutations of the observed compositional vectors across the \(n\) spatial units and evaluating \(I_a\) for each relabeling, which yields the full randomization distribution under the null hypothesis of spatial randomness. For moderate or large \(n\), exhaustive enumeration is infeasible, and the same distribution can be accurately approximated by randomly sampling a large number of permutations and computing \(I_a\) for each draw. This Monte Carlo approximation provides an estimate of the randomization distribution and of derived quantities such as moments and critical values.

\subsection{Exact distribution of \texorpdfstring{$I_a$}{Ia}}

To obtain the exact distribution of $I_a$ under the randomization assumption for a region with \(n\) subareas, the following stjpeg are required:
\begin{enumerate}
    \item Enumerate all \(n!\) permutations of the observed values across the \(n\) subareas.
    \item Compute \(I_a\) for each permutation.
    \item Using the resulting values, compute the desired quantities, such as the expectation, variance, probabilities, and related summaries.
\end{enumerate}

Computing the exact distribution requires evaluating all \(n!\) permutations, which becomes computationally prohibitive even for moderate sample sizes. For example, \(n=9\) already requires \(9!=362{,}880\) evaluations, while \(n=20\) requires more than \(2.4\times 10^{18}\). Therefore, a computationally efficient approximation, such as the Monte Carlo method, is essential.

\subsection{Approximate distribution of \texorpdfstring{$I_a$}{Ia}}

Approximating the distribution using the Monte Carlo method follows the same logic as the exact approach. However, instead of evaluating all \(n!\) permutations, we draw a random sample of permutations and proceed as follows:
\begin{enumerate}
    \item Select \(n^{*}\) permutations from the set of all \(n!\) possible permutations of the observed values across the \(n\) subareas.
    \item Compute \(I_a\) for each of the \(n^{*}\) selected permutations.
    \item Using these values, compute the required quantities, such as the expectation, variance, probabilities, and related summaries.
\end{enumerate}

\subsection{Illustration: \texorpdfstring{$p$}{p} values}

Let \(\pi\) be a permutation of \(\{1,\ldots,n\}\), and define the permuted sample by \(\boldsymbol{x}_{\pi(i)}\), with the same weight matrix \(\mathbf{W}\). The randomization distribution of \(I_a\) is the collection \(\{I_a^{(\pi)}\}\) obtained by evaluating the statistic on each relabeling. This distribution provides a direct basis for inference under the null hypothesis of spatial randomness, including permutation based \(p\) values and critical values.

For the exact distribution, a one sided \(p\) value for positive spatial autocorrelation is
\[
p_{\text{R}}^{+}=\frac{1}{n!}\sum_{\pi}\mathbb{I}\!\left(I_a^{(\pi)}\geq I_a^{\text{obs}}\right),
\]
where \(\mathbb{I}(\cdot)\) denotes the indicator function. Analogously, a one sided \(p\) value for negative autocorrelation is obtained by replacing \(\geq\) with \(\leq\). A two sided \(p\) value can be computed from the randomization distribution using \(|I_a^{(\pi)}|\), or by doubling the smaller one sided \(p\) value, truncated at \(1\). Critical values at level \(\alpha\) are obtained from the corresponding empirical quantiles of \(\{I_a^{(\pi)}\}\).

For the Monte Carlo approximation, let \(B=n^{*}\) and let \(\pi_1,\ldots,\pi_B\) be independent random permutations. A Monte Carlo estimator of the one sided \(p\) value for positive autocorrelation is
\[
\hat{p}_{\text{R}}^{+}=\frac{1}{B}\sum_{b=1}^{B}\mathbb{I}\!\left(I_a^{(\pi_b)}\geq I_a^{\text{obs}}\right),
\]
with analogous definitions for \(\hat{p}_{\text{R}}^{-}\) and for the two sided case. Its Monte Carlo uncertainty can be summarized by the binomial standard error \(\sqrt{\hat{p}(1-\hat{p})/B}\), which makes the dependence on \(B\) explicit.

\section{Spatial correlation for a composition related to the severity of Covid 19 infection in Colombia during January 2021}

Since late 2019, in Wuhan, China, the first cases of pneumonia caused by the SARS CoV 2 virus were reported. It was also established that transmission occurs from person to person and that the rate of spread is high \citet{li2020}. The epidemic of the disease caused by this virus, COVID 19, rapidly expanded across continents, and in March 2020 it was declared a pandemic by the World Health Organization \citep{OMS1}. In Colombia, the first case was confirmed on March 6, 2020, and by the end of 2021 more than five million cases had been confirmed \citep{INS1}.

COVID 19 has a highly heterogeneous clinical course, ranging from asymptomatic infection to severe disease requiring care in an intensive care unit (ICU), and it can also result in death. At the beginning of the pandemic, rapid spread and the absence of population immunity placed substantial pressure on health systems, particularly on ICUs. For this reason, it is relevant to analyze the spatial correlation of the composition of active cases by care setting. If spatial correlation exists and is positive, it supports considering similar actions for a given area and its neighbors. It also suggests potential transmission patterns linked to population mobility, or similarities in population composition, health systems, or baseline health status.

Information on active COVID 19 cases was obtained from the official website of the \textit{Instituto Nacional de Salud} \citep{INS1}. The data are available as one file per day and include variables such as notification date, notification location, including department and municipality, care setting, type of exposure, symptom onset date, diagnosis date, diagnostic test type, active or recovered status, and recovery date, among others.

For each day in January 2021 and for each department, we defined the composition \(\boldsymbol{X}_{ij}=\mathcal{C}(X_{1ij},X_{2ij},X_{3ij})\), where \(X_{1ij}\) is the number of active patients managed at home in department \(i\) on day \(j\), \(X_{2ij}\) is the number of active patients hospitalized in department \(i\) on day \(j\), and \(X_{3ij}\) is the number of active patients in an intensive care unit (ICU) in department \(i\) on day \(j\), with \(j\in\{1,2,3,5,\ldots,31\}\), since data for January 4 were not available.

All computations were carried out in \texttt{R} \citep{R}. Zeros were replaced using the Bayesian Multiplicative replacement method of \citet{fernandez2024}, implemented in the \texttt{zCompositions} package \citep{zCompositions}. Inner products and norms were computed using the \texttt{compositions} package \citep{Rcompositions}. The proposed indicator in (\ref{defIa}) and the Monte Carlo approximation of its distribution in Section~\ref{subsec:distIa} were implemented using custom functions. All code to reproduce our results is publicly available on GitHub.

\begin{figure}[!htb]
\begin{center}
\begin{tabular}{ccc}
  \includegraphics[scale=1]{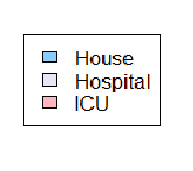} &
  \includegraphics[scale=0.46]{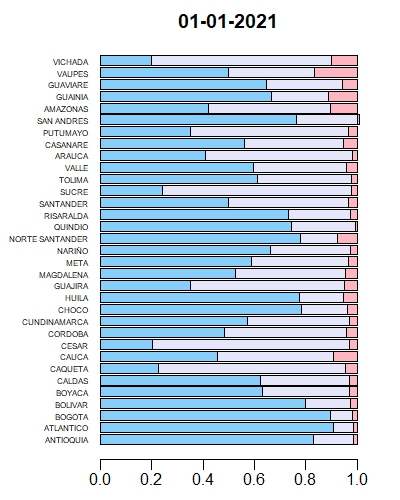} &
  \includegraphics[scale=0.46]{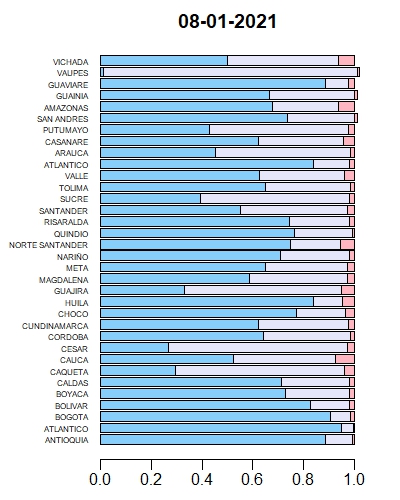}\\
  \includegraphics[scale=0.46]{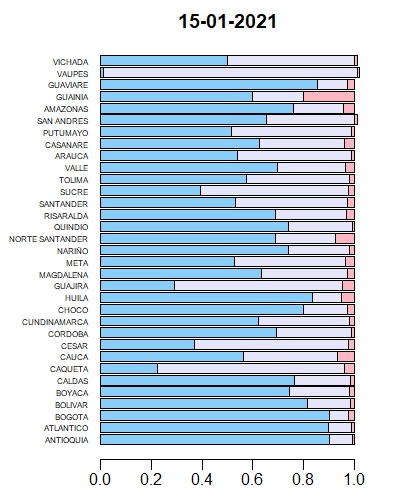} &
  \includegraphics[scale=0.46]{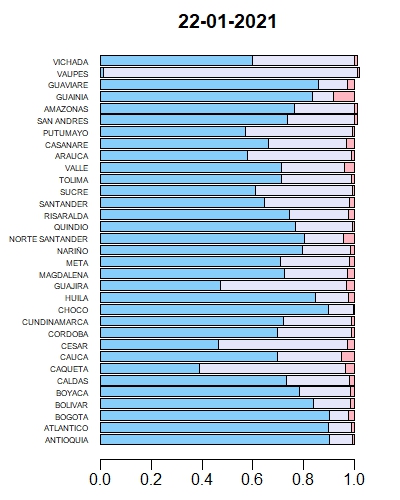} &
  \includegraphics[scale=0.46]{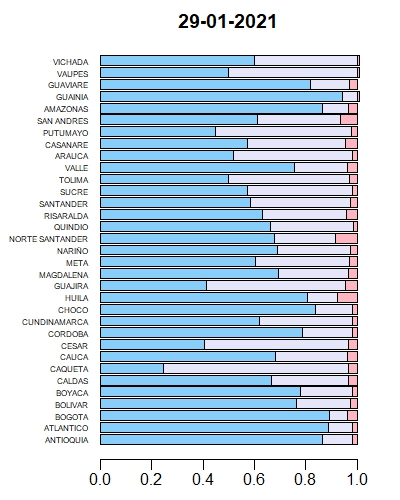} \\
\end{tabular}
\caption{Observed composition by department for selected dates.}\label{comp}
\end{center}
\end{figure}

Overall, across departments and throughout the study period, most active cases were managed at home, with a few exceptions, including \textit{Caquetá}, \textit{Cesar}, \textit{Guajira}, \textit{Sucre}, \textit{Putumayo}, and \textit{Vichada}, which also reported the lowest numbers of confirmed cases. The composition varied over the month. For example, the proportion of individuals requiring specialized care, hospitalization or ICU care, was higher at the beginning of the month. The proportion of active cases admitted to an ICU was low across days and departments. On some days, in the least populated departments, there were no active ICU patients. We also observed zeros in \textit{Vaupés} for the home component and zeros for the ICU component in several departments during the month (Figure~\ref{comp}).

\begin{figure}[!htb]
\begin{center}
\begin{tabular}{cc}
    & 01-01-2021\\
  \includegraphics[scale=1]{leyenda.png} &
  \includegraphics[scale=0.55]{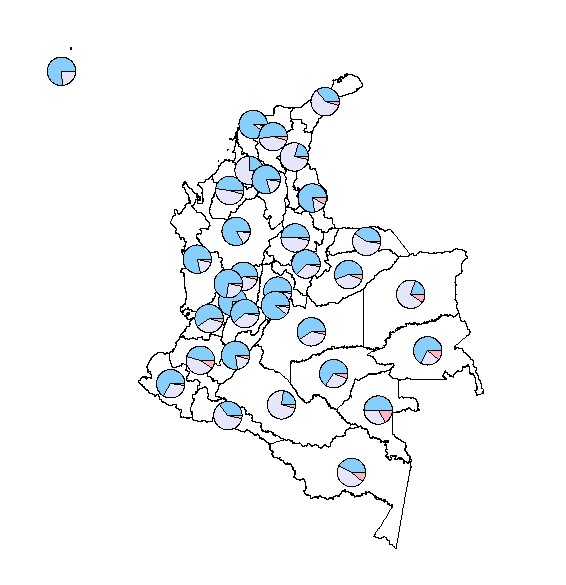}\\
   08-01-2021 & 15-01-2021\\
  \includegraphics[scale=0.55]{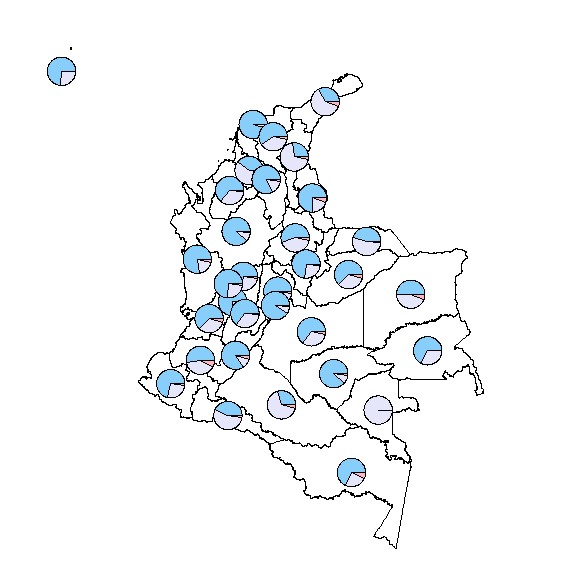} &
  \includegraphics[scale=0.55]{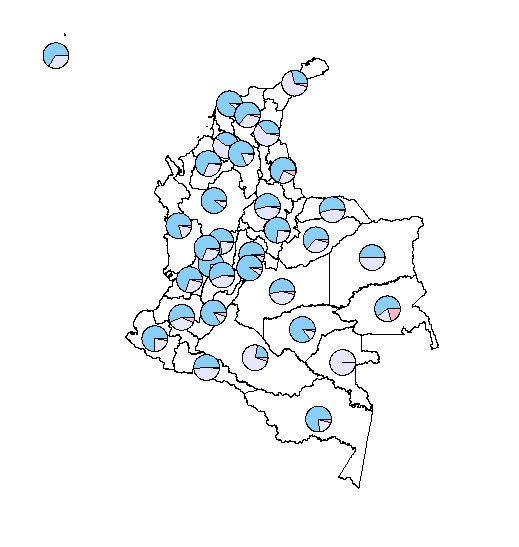}\\
   22-01-2021 & 29-01-2021\\
  \includegraphics[scale=0.55]{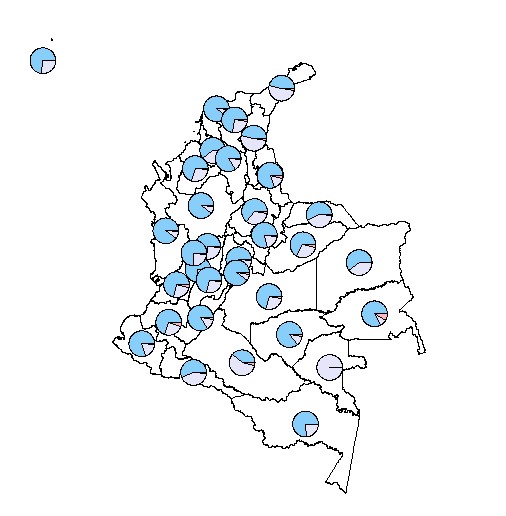} &
  \includegraphics[scale=0.55]{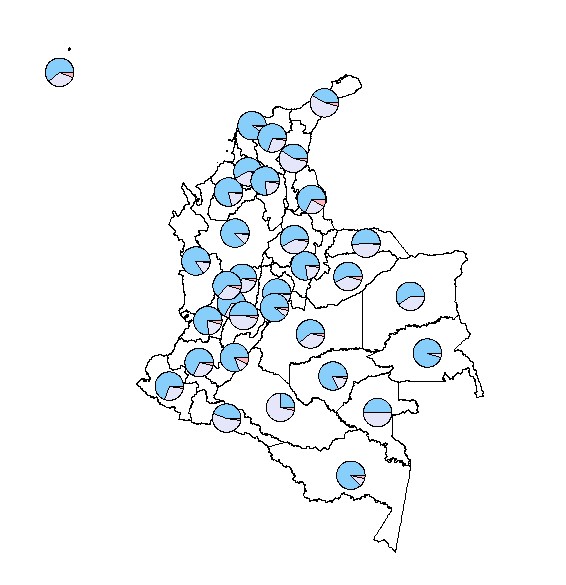} \\
\end{tabular}
\caption{Observed compositions by department for selected dates.}\label{map_comp}
\end{center}
\end{figure}

In addition, Figure~\ref{map_comp} shows that the spatial distribution of the composition changes over the study period. Early in the month, neighboring depatments display similar values, and this pattern evolves as time progresses.

\subsection{Computation of Buitragos's I}

\begin{figure}[!htb]
    \centering
    \includegraphics{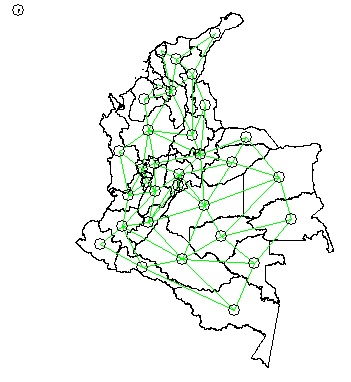}
    \caption{Map of neighborhoods.}
    \label{vecindad}
\end{figure}

Neighborhoods were defined using the queen criterion, so that two departments were considered neighbors if they shared a boundary (Figure~\ref{vecindad}). The corresponding spatial weight matrix was row standardized. Using this matrix, we computed the Reyes's I, \(i_a^j\), as defined in Equation~\ref{defIa}, for each day \(j\in\{1,2,3,5,\ldots,31\}\). We approximated its randomization distribution using the Monte Carlo method described in Section~\ref{subsec:distIa} and estimated the tail probability \(\Pr(I_a>i_a^j)\) as follows:
\begin{enumerate}
    \item Generate \(100{,}000\) spatial permutations of the observed compositions across departments.
    \item Compute the Reyes's I for each permutation.
    \item Estimate \(\hat{p}=\Pr(I_a>i_a^j)\) as the proportion of permutations for which the statistic exceeds the observed value.
\end{enumerate}

\begin{figure}[!htb]
    \centering
    \includegraphics[scale=0.9]{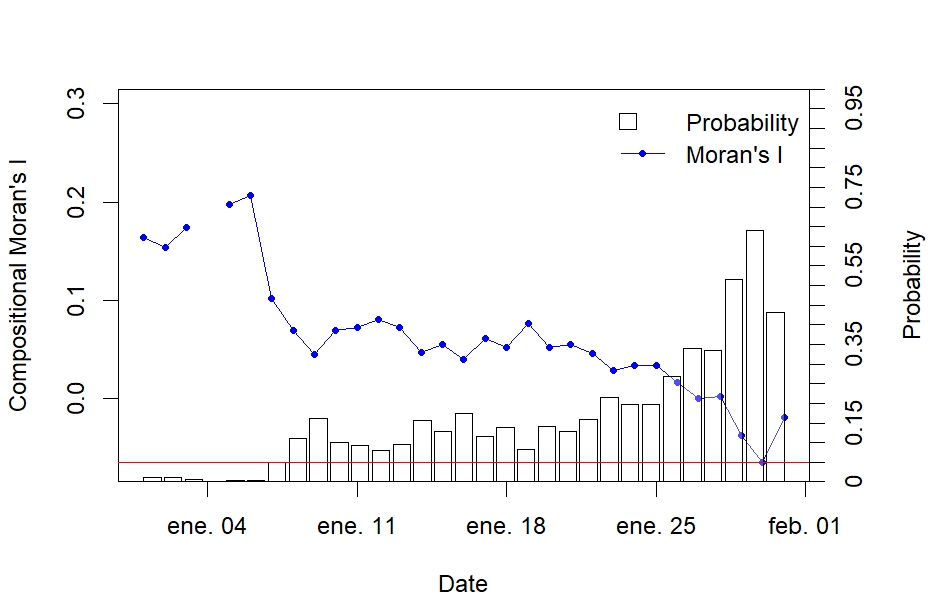}
    \caption{Reyes's I \((i_a^j)\) and $p$ values \(\hat{p}=\Pr(I_a>i_a^j)\).}
    \label{moran_ene}
\end{figure}

Figure~\ref{moran_ene} shows that Reyes's \(I\) decreases as the month progresses. On most days we observed positive spatial autocorrelation for the composition under study. This autocorrelation was statistically significant during the first seven days, reaching its maximum on January 6 \((i_a^6=0.2071,\ \hat{p}=0.0028)\), whereas the smallest positive value was observed on January 28 \((i_a^{28}=0.0023,\ \hat{p}=0.3335)\). The last three days of the month showed negative spatial autocorrelation, although none of these values were statistically significant.

These results suggest that, at the beginning of January 2021, neighboring departments tended to exhibit similar compositions of active cases across home care, hospitalization, and ICU care. In practice, this pattern is consistent with spatial clustering in the severity related distribution of active cases, which may reflect short range transmission dynamics, coordinated health care seeking behavior across adjacent departments, or shared features such as demographic structure, clinical risk profiles, and health system capacity. The progressive decline in Reyes's \(I\) indicates that this spatial structuring weakened over the month, implying increasing heterogeneity across neighboring departments in how active cases were managed. The emergence of negative values near the end of the month, although not statistically significant, points in the direction of local contrasts rather than clustering, where neighboring departments differ more than expected under spatial randomness. From an applied perspective, positive and significant autocorrelation supports the use of spatially informed planning and coordination, since interventions, capacity monitoring, and resource allocation in one department are likely to be relevant for its neighbors, whereas weaker or absent autocorrelation suggests that responses may need to be tailored more locally as spatial coupling diminishes.

\section{Simulation study}

This section presents an extensive simulation study to assess the behavior of Reyes's \(I\) under different scenarios and to compare it with an alternative measure based on the classical Moran's \(I\). Reyes's \(I\) was computed using Equation~\ref{defIa}, and its randomization distribution was approximated via the Monte Carlo method described in Section~\ref{subsec:distIa} using \(n^{*}=10{,}000\) permutations. We defined the alternative indicator as the average of the Moran's \(I\) values computed for each component,
\begin{equation}
    I_m=\frac{1}{D}\sum_{j=1}^D I_j,
\end{equation}\label{Itrad}
where \(I_j\) denotes Moran's \(I\) for component \(j=1,\ldots,D\). Its distribution was also approximated via Monte Carlo as follows:
\begin{enumerate}
    \item For a total of \(n\) subareas, draw \(n^{*}=10{,}000\) permutations.
    \item Compute \(I_m\) for each permutation.
    \item Use the resulting values to compute the required summaries.
\end{enumerate}
We considered three settings: identical compositions across subareas, independent compositions across subareas, and spatially correlated compositions. For each setting, we simulated square lattices of size \(n\times n\), with \(n\in\{3,5,7,10\}\), and compositions with \(D\in\{3,5,7\}\) components. For each case, we constructed the spatial weight matrix using both the queen and rook criteria.

\subsection{Case 1: Identical compositions across all subareas}

In this setting, the common composition was generated from a logistic normal distribution on the simplex with mean \(\mathcal{C}(1,\ldots,1)\) and three covariance structures: identity, exchangeable, and Wishart generated, with a Toeplitz scale matrix. For each of the \(1000\) replications and each covariance structure, we computed the proposed indicator and the upper bound in Equation~\ref{bound}. As expected, since all lattice values are identical, the proposed indicator equals the upper bound in every simulation, attaining the maximum spatial autocorrelation.

\subsection{Case 2: Independent compositional data}

In this case, values were again simulated independently for each subarea from a logistic normal distribution on the simplex, under the three covariance structures described above, identity, exchangeable, and Wishart generated. For each of the \(10{,}000\) replications, we computed the proposed indicator \(I_a\), the alternative indicator \(I_m\), and the processing time required to compute each indicator together with its Monte Carlo approximation of the randomization distribution. We also estimated the tail probability that each indicator exceeds its observed value. For the proposed indicator, we used
\begin{equation}
    \widehat{\Pr}\!\left(I_a>I_{\text{obs}}\right)=\frac{m}{n^{*}},
\end{equation}\label{pvalue}
where \(m\) is the number of Monte Carlo permutations for which \(I_a\) exceeds the observed value \(I_{\text{obs}}\). The same calculation was used for \(I_m\).

\begin{figure}[!htb]
	\centering
	\setlength{\tabcolsep}{0pt}
	\begin{tabular}{ccc}
		& \textsf{Queen} & \textsf{Rook} \\
		\begin{sideways} \hspace{1.7cm} \textsf{Identity} \end{sideways} &
		\includegraphics[scale = 0.45]{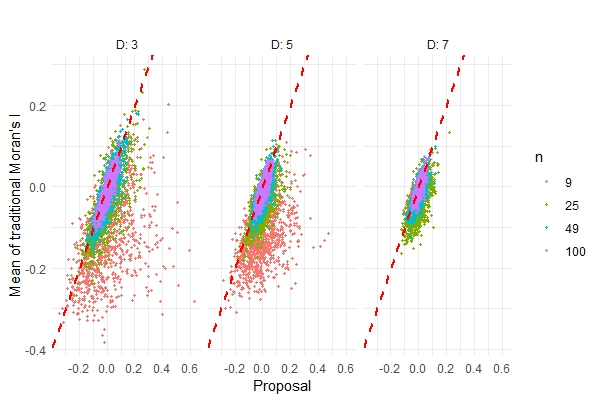} &
		\includegraphics[scale = 0.45]{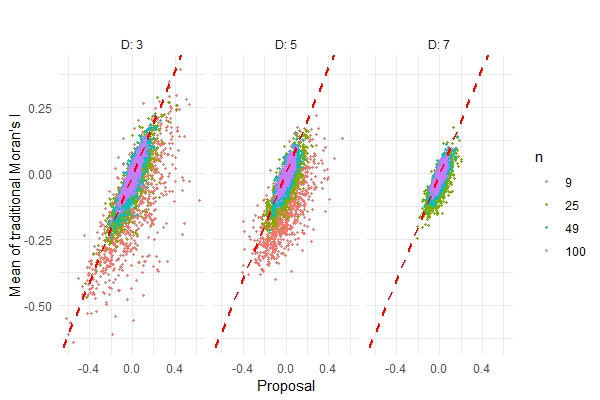} \\
		\begin{sideways} \hspace{1.2cm} \textsf{Exchangeable} \end{sideways} &
		\includegraphics[scale = 0.45]{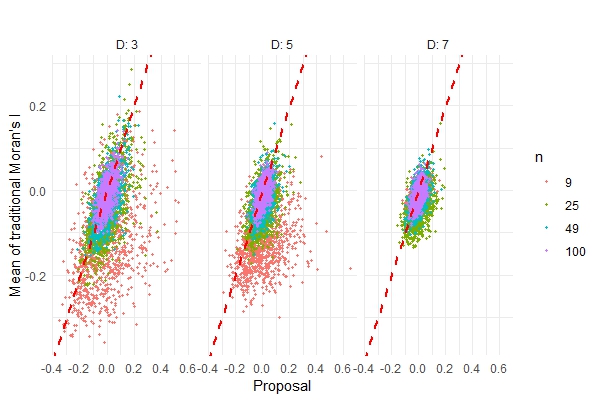} &
		\includegraphics[scale = 0.45]{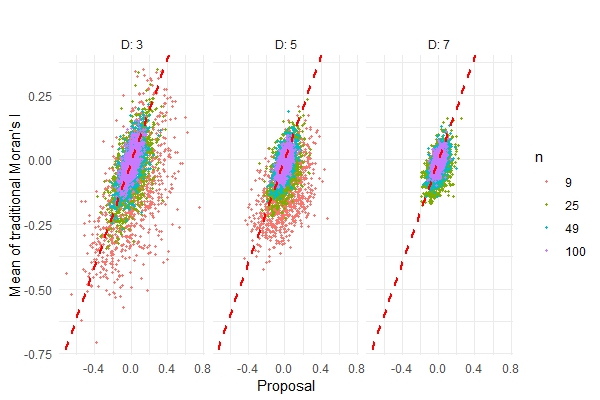} \\
		\begin{sideways} \hspace{1.7cm} \textsf{Wishart} \end{sideways} &
		\includegraphics[scale = 0.45]{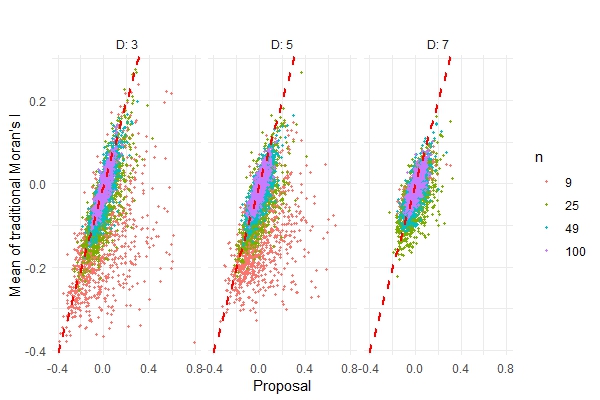} &
		\includegraphics[scale = 0.45]{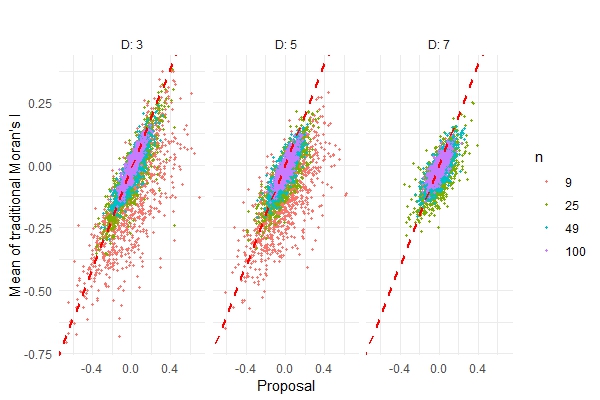} \\
	\end{tabular}
	\caption{Reyes's $I$ versus the alternative indicator $I_m$ under three covariance structures (Identity, Exchangeable, Wishart), comparing spatial weights based on queen and rook contiguity.}
	\label{prop_vs_trad_queen_rook}
\end{figure}

\begin{figure}[!htb]
	\centering
	\setlength{\tabcolsep}{0pt}
	\begin{tabular}{ccc}
		& \textsf{Queen} & \textsf{Rook} \\
		\begin{sideways} \hspace{1.7cm} \textsf{Identity} \end{sideways} &
		\includegraphics[scale = 0.45]{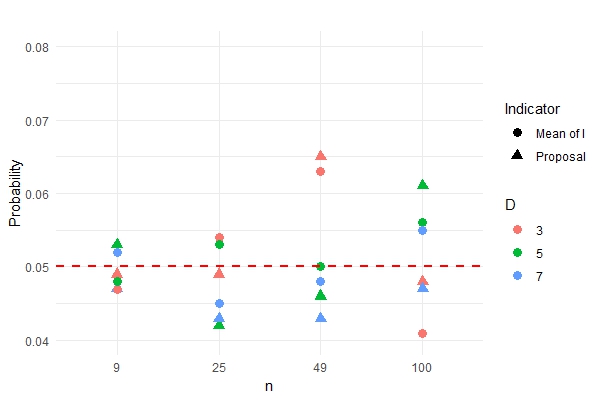} &
		\includegraphics[scale = 0.45]{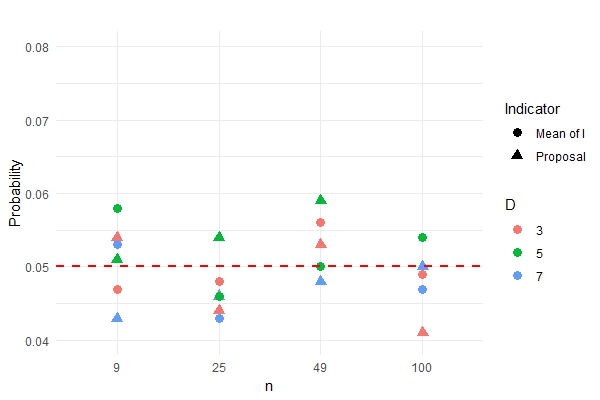} \\
		\begin{sideways} \hspace{1.2cm} \textsf{Exchangeable} \end{sideways} &
		\includegraphics[scale = 0.45]{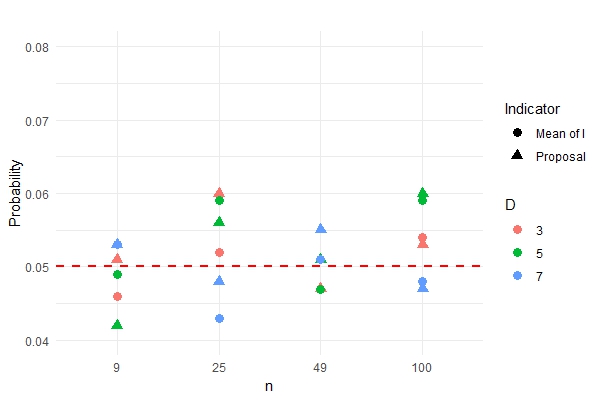} &
		\includegraphics[scale = 0.45]{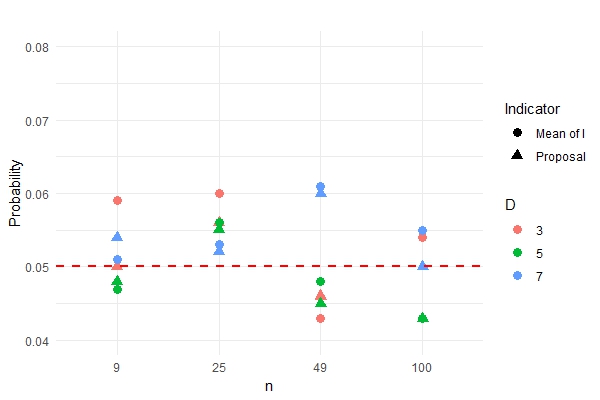} \\
		\begin{sideways} \hspace{1.7cm} \textsf{Wishart} \end{sideways} &
		\includegraphics[scale = 0.45]{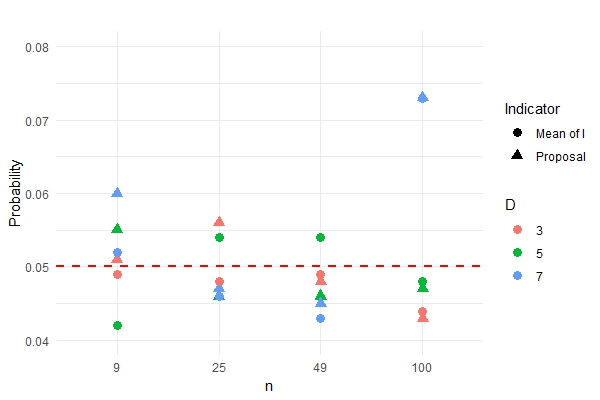} &
		\includegraphics[scale = 0.45]{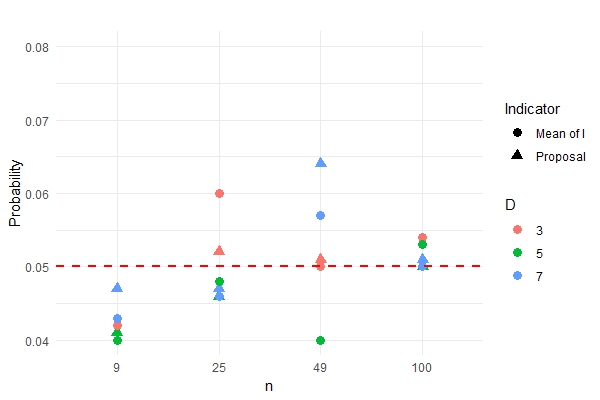} \\
	\end{tabular}
	\caption{Grid size versus type I error probability, by indicator and covariance structure (Identity, Exchangeable, Wishart), comparing spatial weights based on queen and rook contiguity.}
	\label{fig_significance2_queen_rook}
\end{figure}

\begin{figure}[!htb]
	\centering
	\setlength{\tabcolsep}{0pt}
	\begin{tabular}{ccc}
		& \textsf{Queen} & \textsf{Rook} \\
		\begin{sideways} \hspace{1.7cm} \textsf{Identity} \end{sideways} &
		\includegraphics[scale = 0.45]{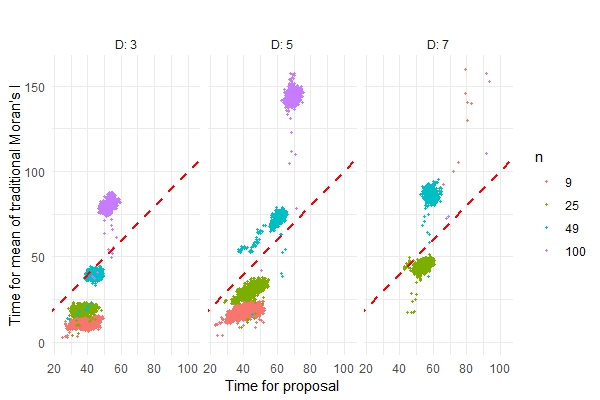} &
		\includegraphics[scale = 0.45]{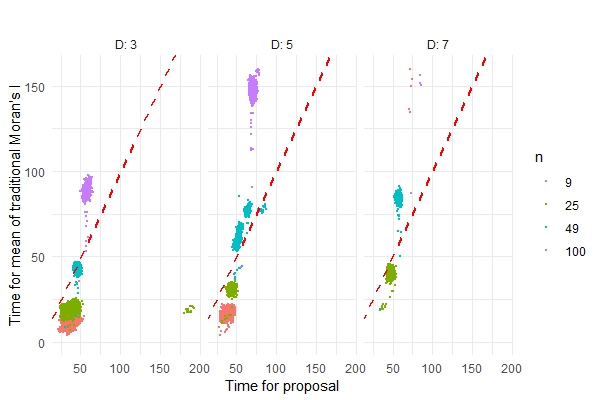} \\
		\begin{sideways} \hspace{1.2cm} \textsf{Exchangeable} \end{sideways} &
		\includegraphics[scale = 0.45]{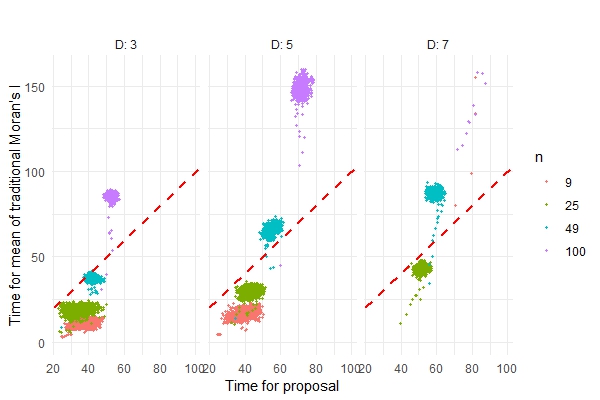} &
		\includegraphics[scale = 0.45]{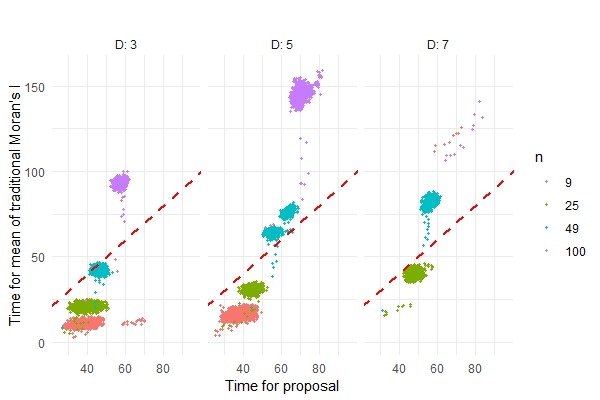} \\
		\begin{sideways} \hspace{1.7cm} \textsf{Wishart} \end{sideways} &
		\includegraphics[scale = 0.45]{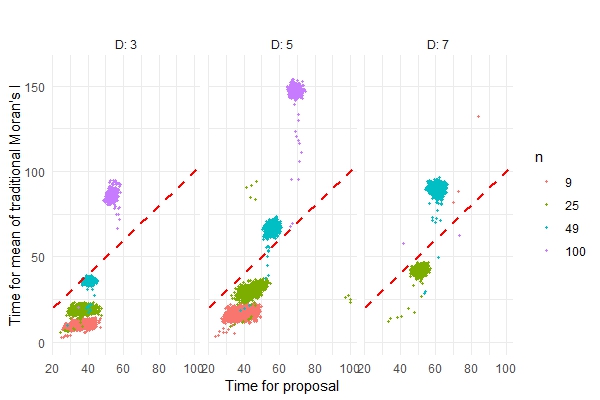} &
		\includegraphics[scale = 0.45]{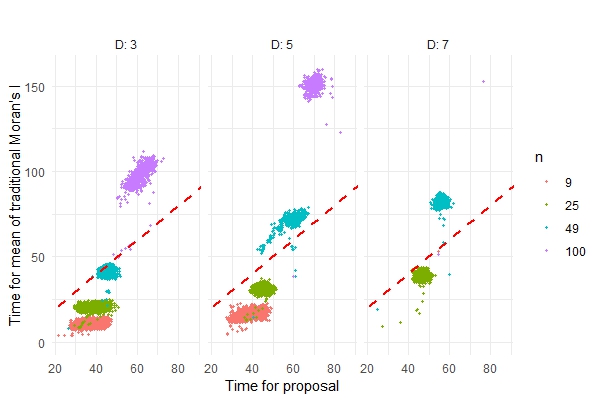} \\
	\end{tabular}
	\caption{Processing times for the proposed indicator versus the alternative indicator under three covariance structures (Identity, Exchangeable, Wishart), comparing spatial weights based on queen and rook contiguity.}
	\label{fig_tiempos2_queen_rook}
\end{figure}

Overall, the two indicators yield similar values, and their agreement increases with grid size. As the number of components increases, the scatter concentrates around the line \(y=x\) and the variability of both indicators decreases, with \(I_m\) consistently more variable than \(I_a\). Across covariance structures, the identity matrix yields the smallest variability, whereas the Wishart based covariance yields the largest. Results under the queen and rook neighborhood criteria are broadly similar (Figure~\ref{prop_vs_trad_queen_rook}).

Taken together, these simulations indicate that, under independence across subareas, Reyes's \(I\) and the componentwise average \(I_m\) yield similar values, with agreement improving as grid size increases, while \(I_m\) remains more variable, consistent with the fact that averaging marginal Moran statistics can propagate component level noise relative to a single compositional measure defined in Aitchison geometry. As the number of components grows, both indicators become less variable and concentrate more tightly around the line \(y=x\), and results are broadly insensitive to whether queen or rook neighborhoods are used.

To assess inferential calibration, we approximated the probability of rejecting the null hypothesis of no spatial autocorrelation when it is true as the proportion of simulations with \(\widehat{\Pr}(I_a>I_{\text{obs}})<0.05\). These estimated Type I error rates are generally close to \(0.05\), with no clear pattern across grid sizes or numbers of components. Under the queen criterion, \(I_a\) is closer to the nominal level than \(I_m\) when the covariance is identity and \((D,\text{grid size})\) equals \((3,9)\), \((3,25)\), \((3,100)\), \((5,25)\), \((7,25)\), \((7,49)\), or \((7,100)\) (Figure~\ref{fig_significance2_queen_rook}). Under the rook criterion, \(I_a\) is closer to \(0.05\) than \(I_m\) when the covariance is identity and \((D,\text{grid size})\) equals \((3,49)\), \((5,9)\), \((5,100)\), \((7,49)\), or \((7,100)\), when the covariance is exchangeable and \((D,\text{grid size})\) equals \((3,9)\), \((3,49)\), \((5,9)\), \((5,25)\), \((7,25)\), or \((7,100)\), and when the covariance is Wishart based and \((D,\text{grid size})\) equals \((3,49)\), \((5,100)\), \((7,9)\), \((7,25)\), or \((7,100)\) (Figure~\ref{fig_significance2_queen_rook}).

Regarding processing time, runtimes increase with grid size and with the number of components for both indicators. For \(D=3\), \(I_a\) is faster than \(I_m\) except on \(10\times 10\) grids, whereas for \(D\in\{5,7\}\), \(I_a\) is faster on \(3\times 3\) and \(5\times 5\) grids. Processing times are broadly similar across covariance structures (Figure~\ref{fig_tiempos2_queen_rook}), suggesting that the main drivers of computational cost are lattice size, dimension, and the Monte Carlo procedure rather than the specific covariance specification.

\subsection{Case 3: Spatially correlated compositional data}

In this case, we simulated spatially correlated data in \(\mathbb{R}^{D-1}\) using a multivariate spatial autoregressive model,
\begin{equation}
    \mathbf{Y}^*=\left(\mathbf{I}-\rho \mathbf{W}\right)^{-1}\mathbf{E}^*, \label{SAR}
\end{equation}
with \(\rho\in\{0.5,0.7,0.9\}\), as follows:
\begin{enumerate}
    \item Generate \(\mathbf{E}^*=(\boldsymbol{\epsilon}_1^*,\ldots,\boldsymbol{\epsilon}_n^*)\top \in \mathbb{R}^{n\times(D-1)}\), where the vectors \(\boldsymbol{\epsilon}_i^*\in\mathbb{R}^{D-1}\) are independent and satisfy
    \(\boldsymbol{\epsilon}_i^*\sim \textsf{N}\!\left(\boldsymbol{0}_{D-1},\mathbf{\Sigma}\right)\), with \(\mathbf{\Sigma}\) given by one of the following covariance structures:
    \begin{itemize}
        \item Identity.
        \item Exchangeable, \(\Sigma_{ii}=1\) and \(\Sigma_{ij}=\rho_1\), with \(-1/(D-2)\leq \rho_1 \leq 1\) to ensure positive definiteness.
        \item Wishart generated, with a Toeplitz scale matrix.
    \end{itemize}
    \item Compute \(\mathbf{Y}^*\) using Equation~\ref{SAR}.
    \item Apply \(\operatorname{ilr}^{-1}\) rowwise to map the resulting values from \(\mathbb{R}^{D-1}\) to \(S^{D}\).
\end{enumerate}

For each combination of \(\rho\) and covariance structure, we generated \(1000\) replications. In each replication, we computed the proposed indicator \(I_a\), the alternative indicator \(I_m\), and the processing time required to compute each indicator together with its Monte Carlo approximation of the randomization distribution. As in Case~2, we also estimated the tail probability that each indicator exceeds its observed value using Equation~\ref{pvalue}.

\begin{figure}[!htb]
	\centering
	\setlength{\tabcolsep}{0pt}
	\begin{tabular}{ccc}
		& \textsf{Queen} & \textsf{Rook} \\
		\begin{sideways} \hspace{1.7cm} \textsf{Identity} \end{sideways} &
		\includegraphics[scale = 0.45]{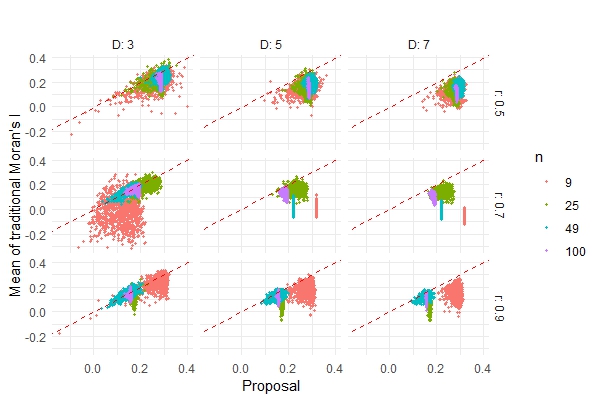} &
		\includegraphics[scale = 0.45]{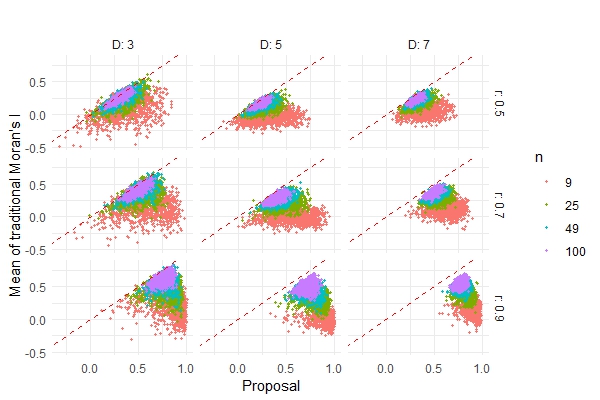} \\
		\begin{sideways} \hspace{1.2cm} \textsf{Exchangeable} \end{sideways} &
		\includegraphics[scale = 0.45]{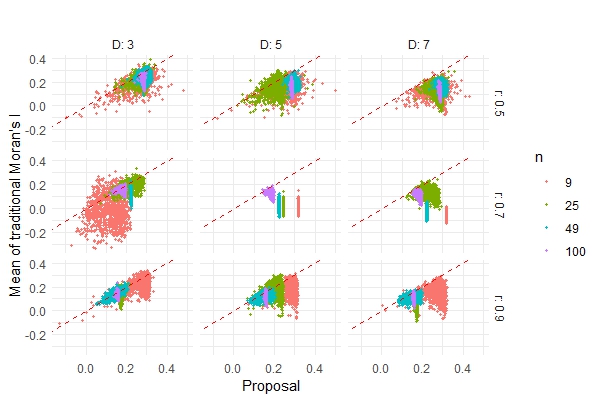} &
		\includegraphics[scale = 0.45]{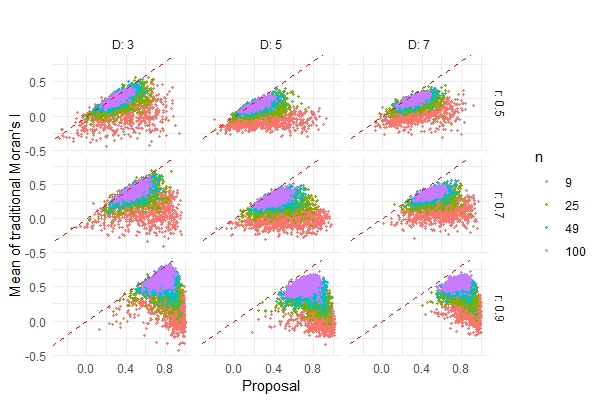} \\
		\begin{sideways} \hspace{1.7cm} \textsf{Wishart} \end{sideways} &
		\includegraphics[scale = 0.45]{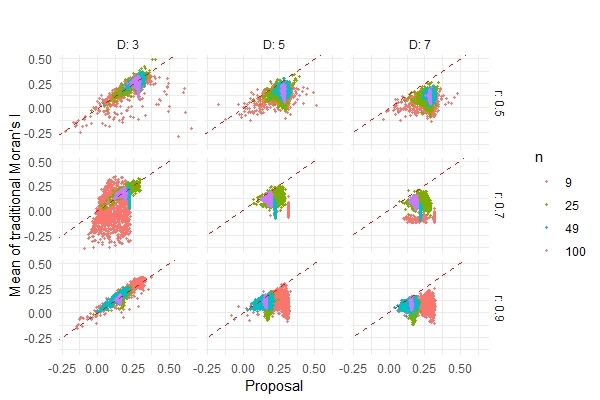} &
		\includegraphics[scale = 0.45]{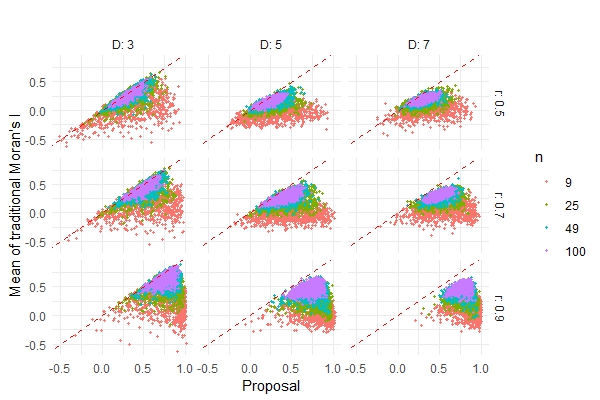} \\
	\end{tabular}
	\caption{Reyes's \(I\) versus the alternative indicator \(I_m\) under spatial dependence for three covariance structures (Identity, Exchangeable, Wishart), comparing spatial weights based on queen and rook contiguity.}
	\label{prop_vs_trad3_queen_rook}
\end{figure}

\begin{figure}[!htb]
	\centering
	\setlength{\tabcolsep}{0pt}
	\begin{tabular}{ccc}
		& \textsf{Queen} & \textsf{Rook} \\
		\begin{sideways} \hspace{1.7cm} \textsf{Identity} \end{sideways} &
		\includegraphics[scale = 0.45]{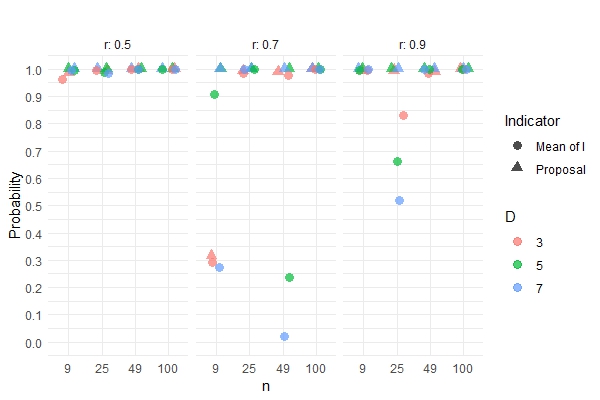} &
		\includegraphics[scale = 0.45]{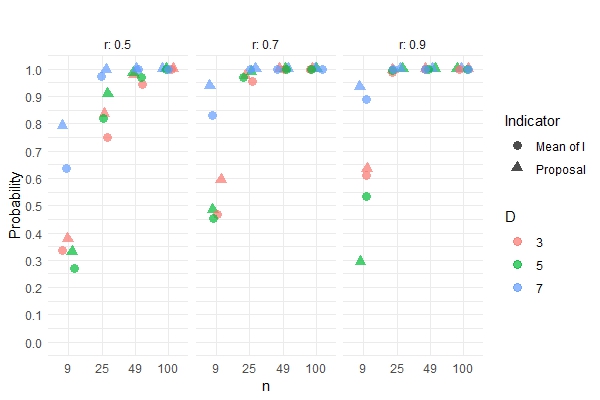} \\
		\begin{sideways} \hspace{1.2cm} \textsf{Exchangeable} \end{sideways} &
		\includegraphics[scale = 0.45]{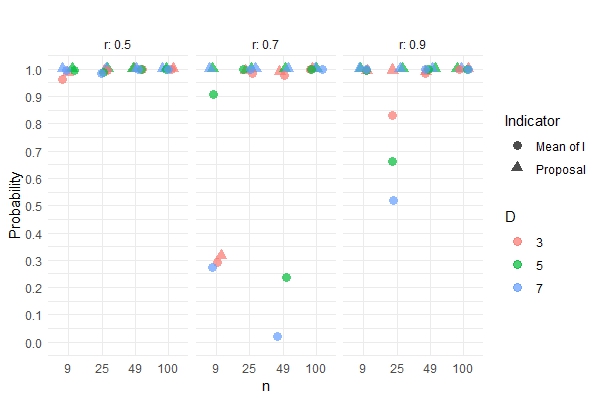} &
		\includegraphics[scale = 0.45]{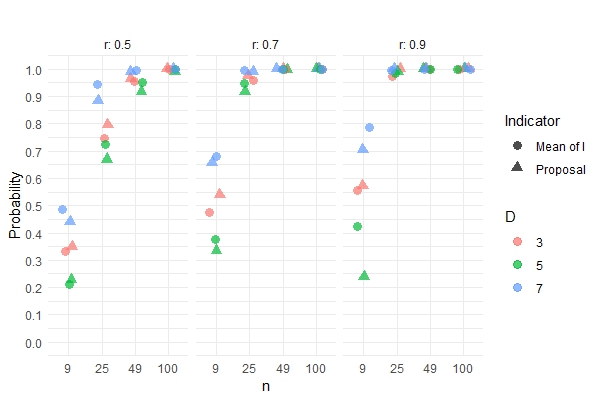} \\
		\begin{sideways} \hspace{1.7cm} \textsf{Wishart} \end{sideways} &
		\includegraphics[scale = 0.45]{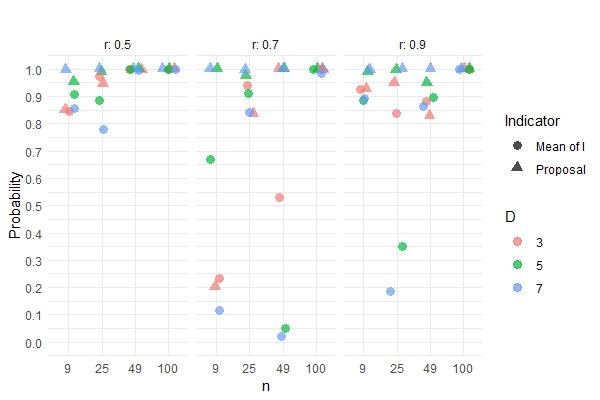} &
		\includegraphics[scale = 0.45]{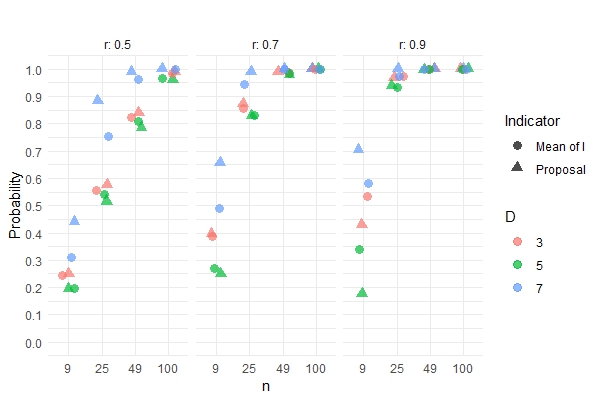} \\
	\end{tabular}
	\caption{Grid size versus empirical rejection rate, by indicator and covariance structure (Identity, Exchangeable, Wishart), comparing spatial weights based on queen and rook contiguity.}
	\label{fig_probability3_queen_rook}
\end{figure}

\begin{figure}[!htb]
	\centering
	\setlength{\tabcolsep}{0pt}
	\begin{tabular}{ccc}
		& \textsf{Queen} & \textsf{Rook} \\
		\begin{sideways} \hspace{1.7cm} \textsf{Identity} \end{sideways} &
		\includegraphics[scale = 0.45]{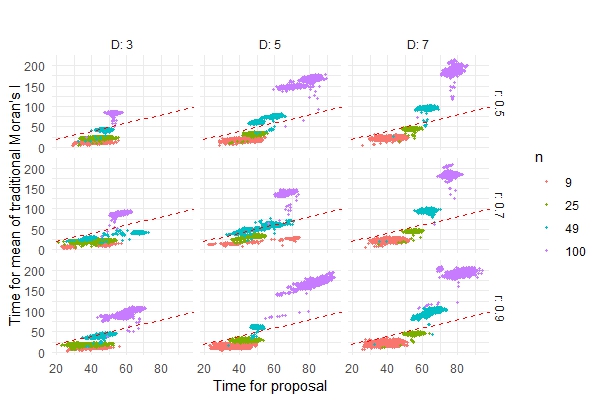} &
		\includegraphics[scale = 0.45]{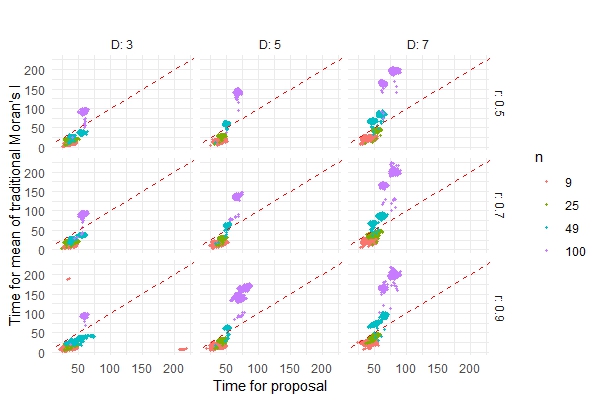} \\
		\begin{sideways} \hspace{1.2cm} \textsf{Exchangeable} \end{sideways} &
		\includegraphics[scale = 0.45]{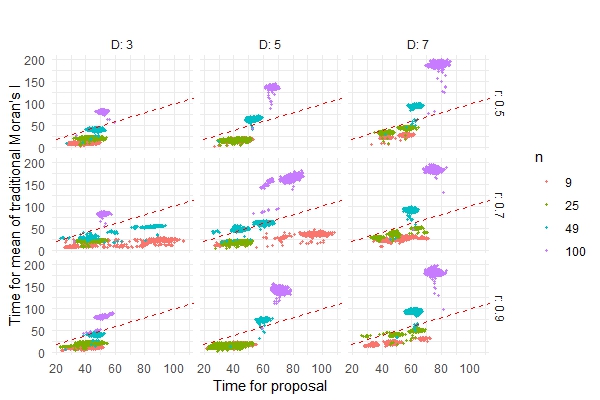} &
		\includegraphics[scale = 0.45]{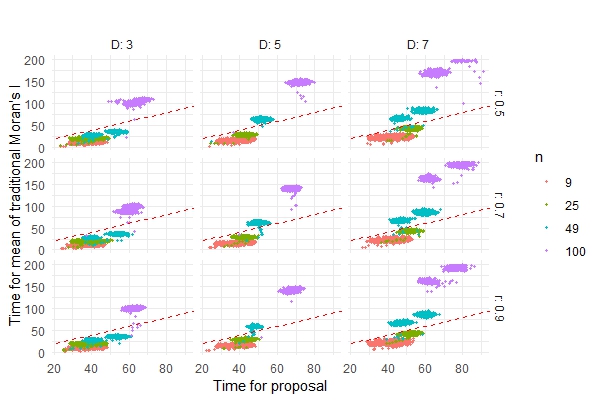} \\
		\begin{sideways} \hspace{1.7cm} \textsf{Wishart} \end{sideways} &
		\includegraphics[scale = 0.45]{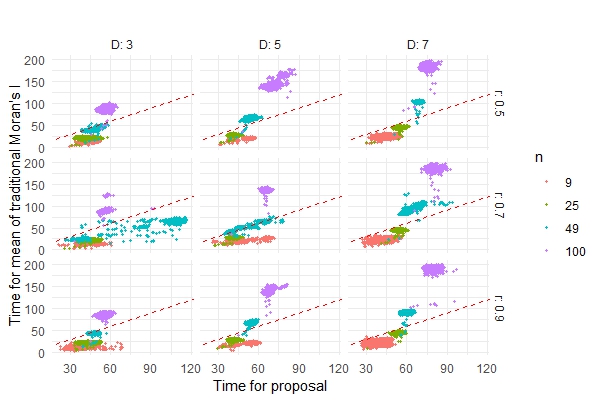} &
		\includegraphics[scale = 0.45]{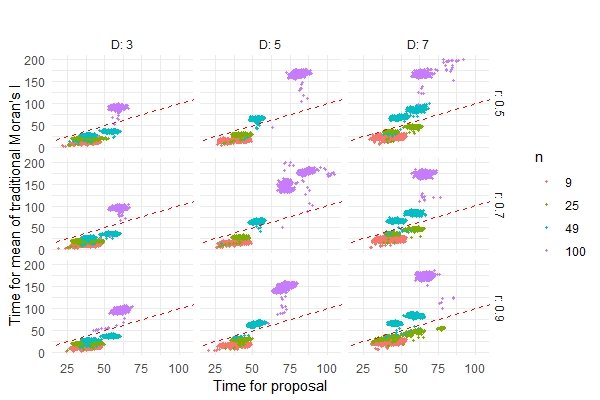} \\
	\end{tabular}
	\caption{Processing times for the proposed indicator versus the alternative indicator under spatial dependence for three covariance structures (Identity, Exchangeable, Wishart), comparing spatial weights based on queen and rook contiguity.}
	\label{fig_tiempos3_queen_rook}
\end{figure}

Overall, under both neighborhood criteria we typically observe \(I_a<I_m\). As the number of components decreases, the two indicators become closer on average, but their variability increases, with the largest dispersion occurring on small lattices, especially for \(n=9\) and \(\rho=0.7\). In several configurations, again mainly at \(\rho=0.7\), \(I_a\) shows markedly lower variability than \(I_m\), suggesting that \(I_a\) responds more homogeneously to moderate spatial dependence. These qualitative patterns are stable across covariance structures, and the identity covariance generally yields the smallest dispersion.

To assess evidence against the null hypothesis of no spatial autocorrelation under this data generating mechanism, we computed the empirical rejection rate, defined as the proportion of replications with \(\widehat{\Pr}(I_a>I_{\text{obs}})<0.05\), and analogously for \(I_m\). These rejection rates are typically close to \(1\) for both indicators, indicating high power in most settings, and they do not exhibit a clear monotone pattern in \(D\) or grid size. Under queen contiguity, \(I_m\) often yields smaller rejection rates than \(I_a\) when the covariance is identity or exchangeable, while under the Wishart generated covariance the behavior is more heterogeneous, although rejection rates remain mostly above \(0.8\) and are frequently larger for \(I_a\), with occasional reversals when \(D=3\).

Under rook contiguity, departures from rejection rates near \(1\) are more frequent for both indicators, although \(I_a\) still tends to be larger. The smallest rejection rates occur primarily on the smallest lattices, and the deviation from \(1\) becomes more pronounced as \(\rho\) decreases, consistent with weaker spatial signal. Under the Wishart generated covariance, \(I_a\) can be smaller than \(I_m\) in some small sample configurations, including \(n=9\) with \(D=5\) across values of \(\rho\), and also for \(\rho=0.9\) with \(D=3\).

Regarding processing time, runtimes increase with grid size and with the number of components, as in Case~2. For larger lattices, computing \(I_a\) is generally faster than computing \(I_m\), and differences across covariance structures are modest, indicating that computational cost is driven mainly by lattice size, dimension, and the Monte Carlo procedure.

\section*{Discussion}

In this study we introduce a new measure of spatial autocorrelation tailored to areal compositional data. The proposed indicator is built from the Aitchison norm and inner product, which are the natural geometric objects on the simplex, the sample space of compositions. This choice ensures mathematical coherence with the data structure while preserving key properties required for compositional analysis, including scale invariance and invariance to permutations of the parts. In addition, the indicator is invariant to the choice of contrast matrix in the \(\operatorname{ilr}\) transformation, which strengthens robustness and facilitates implementation.

Regarding theoretical properties, we established an upper bound, the randomization expected value, and the noncentral second moment under the randomization assumption. The expected value coincides with that of the classical Moran index for real valued data, providing a clear link between the proposed indicator and standard spatial autocorrelation theory. Although the noncentral second moment is not identical to that of the conventional Moran statistic, its form is consistent with previously known results for Moran, and it yields a useful characterization of the variability of the proposed indicator under the null.

The simulation study, across three data generating scenarios, is broadly consistent with these theoretical results. When all subareas share the same composition, the proposed indicator \(I_a\) attains the theoretical bound exactly, confirming its behavior in an extreme configuration. Under independence across subareas, \(I_a\) and the alternative \(I_m\) produce similar values and their agreement increases with grid size, but \(I_a\) is typically less variable, especially as the number of components grows, and it tends to track the nominal significance level more closely. When data are generated with explicit spatial autocorrelation, both indicators consistently detect dependence, with rejection probabilities close to \(1\) in most configurations, but \(I_a\) often exhibits more homogeneous sampling behavior and reduced sensitivity to changes in the covariance structure, suggesting greater stability under complex dependence. From a computational perspective, runtimes increase with both grid size and compositional dimension for the two procedures, yet \(I_a\) is generally more efficient on larger lattices, which is particularly relevant when permutation based inference requires many repetitions.

Taken together, these findings indicate that, while differences between \(I_a\) and \(I_m\) are not uniformly large, the proposed indicator offers systematic advantages in stability, control of the nominal level, and computational efficiency. These features make \(I_a\) a competitive and methodologically appropriate option for assessing spatial autocorrelation in compositional settings.

The application to COVID-19 infection severity in Colombia during January 2021 illustrates practical relevance. The results show positive and significant spatial autocorrelation early in the month, followed by a gradual attenuation to nonsignificance toward the end of the period. This pattern reflects a changing spatial structure in the pandemic and highlights the ability of the indicator to capture temporal variation in spatial dependence for compositional outcomes such as home care, hospitalization, and ICU. The fact that \(I_a\) can be applied directly to such compositions, while respecting the geometry of the simplex, supports its use in epidemiological and public health analyses.

Several directions remain for future work. First, it would be valuable to study the indicator under alternative distributions on the simplex beyond the logistic normal, to assess whether the similarities observed with \(I_m\) are driven by the normality assumption adopted here. Second, further exploration across a wider range of spatial dependence parameters within SAR type mechanisms would extend and refine the results of the third simulation scenario. Finally, evaluating performance under spatial models beyond SAR would help establish robustness and applicability in settings with more complex spatial dynamics.

\section*{Statements and declarations}

The authors declare that they have no known competing financial interests or personal relationships that could have appeared to influence the work reported in this article.

All R and C++ code required to reproduce our results is publicly available at {\footnotesize \url{https://github.com/labuitragor-arch/Measuring-Spatial-Autocorrelation-in-Compositions}}. The re\-po\-si\-to\-ry includes a detailed README with step by step instructions, and the scripts are well documented. All datasets used in the applications and cross validation exercises are also included in the repository.

During the preparation of this work the authors used ChatGPT-5-Thinking in order to improve language and readability. After using this tool, the authors reviewed and edited the content as needed and take full responsibility for the content of the publication.

%\nocite{*}
\bibliography{references.bib}
\bibliographystyle{apalike}

% \appendix

\end{document}